\documentclass[%
reprint,10pt,
nofootinbib,
nobibnotes,
 amsmath,amssymb,
pra,
onecolumn,notitlepage]{revtex4-1}
\usepackage{amsmath,array}
\usepackage{multirow}
\usepackage{tikz}
\usepackage{graphicx}% Include figure files
\usepackage{color}
\usepackage{amssymb}
\usepackage{mathtools}
\usepackage{dcolumn}% Align table columns on decimal point
\usepackage{bm}% bold math
\usepackage{epstopdf}
\AppendGraphicsExtensions{.tif}
\begin{document}

%\preprint{On the families  of Wigner functions }

\title[On  families  of Wigner functions]{On families of Wigner functions for {\boldmath$N$}-level quantum systems}% Force line breaks with \\
%%\thanks{A footnote to the article title}%

\author{Vahagn Abgaryan}%
 \email{vahagnab@googlemail.com}\affiliation{Laboratory of Information Technologies, Joint Institute for Nuclear Research, Dubna,  Russia}
 \affiliation{A. I. Alikhanyan National Science Laboratory, 0036 Yerevan, Armenia}
 \affiliation{Department of Mathematics, Czech Technical University,
Prague 12000, Czech Republic}
 
\author{Arsen Khvedelidze}
 \email{akhved@jinr.ru.}%Lines break automatically or can be forced with \\
\affiliation{Laboratory of Information Technologies, Joint Institute for Nuclear Research, Dubna, Russia}
\affiliation{A. Razmadze Mathematical Institute, Iv.Javakhishvili Tbilisi State University,      Tbilisi, Georgia}
\affiliation{Institute of Quantum Physics and Engineering Technologies, Georgian Technical University, Tbilisi, Georgia}

\begin{abstract}
Based on the Stratonovich-Weyl correspondence, 
a method  of constructing the unitary non-equivalent Wigner quasiprobability distributions  for a  generic $N$-level quantum system is proposed. 
The mapping between the operators on the Hilbert space and the functions on the phase space is implemented by the Stratonovich-Weyl operator kernel.
The algebraic ``master equation'' for the Stratonovich-Weyl kernel is derived  and the ambiguity in its solution is analyzed.
The general method is exemplified by considering the Wigner functions of  a single qubit and a single qutrit. 
\end{abstract}
%\pacs{AAAAAAAAAAAAAA}% PACS, the Physics and Astronomy
                             % Classification Scheme.
%\keywords{Suggested keywords}%Use showkeys class option if keyword
                              %display desired
\maketitle
%\tableofcontents
\section{Introduction}
\label{intro}
A modern boom in quantum engineering and quantum computing gave new life to the studies of an interplay between classical and quantum physics. 
Particularly, a new insight has been gained into the 
long-standing problem of finding ``quantum analogues'' for the statistical distributions of classical systems.
The Wigner procedure~\cite{Wigner1932} to associate 
the so-called ``quasiprobability distribution'' on a phase space with a density operator acting on a Hilbert space was essentially the definition of the inverse of the 
Weyl  quantization rule~\cite{Weyl1928}. The discovery of  this mapping  provided the formulation of one of the most interesting representations of the quantum theory as a statistical theory on a phase
space, which is usually connected to the names of Groenewold~\cite{Groenewold1946} and Moyal \cite{Moyal1949}. 
After almost a century of elaboration of the initial  ideas,  diverse aspects of the  interrelations between the phase space functions and the operators on  the Hilbert space have been established (e.g. Refs.~ \onlinecite{HilleryOConnellScullyWigner1984} -\onlinecite{BrifMann1999}). 
Nowadays, as it was mentioned in the beginning of the article, special attention is drawn due to quantum engineering needs,  to the considerations of the phase-space formulation of the  quantum theory including the studies of the Wigner quasiprobability distributions for finite-dimensional quantum systems (cf. Ref.~\onlinecite{TilmaEverittSamsonMunroNemoto2016}  and references therein).  

In the present note we continue these studies and
discuss the issue of the non-uniqueness of the mapping between quantum and classical descriptions.
Based on the  postulates known as the Stratonovich-Weyl correspondence~\cite{Stratonovich}, a method of determining the Wigner quasiprobability distributions (shortly, the 
Wigner functions (WF)) for a generic $N$-level quantum system is suggested.  
The Wigner function is constructed from two objects: the density matrix $\varrho$ describing a quantum state, and the so-called Stratonovich-Weyl (SW) kernel $\Delta(\Omega_N )$ defined over the symplectic manifold $\Omega_N\,.$ 
As it will be shown below, starting from the first principles,  the  kernel $\Delta(\Omega_N )$ is subject to a set of algebraic equations.  According to those  equations, the SW kernel for a given quantum state $\varrho$  depends on a set of $N-2$ real parameters 
$\boldsymbol{\nu}=( \nu_1, \nu_2, \dots, \nu_{N-2})\,.$
Moreover, these SW kernels 
$\Delta(\Omega_N\,|\,\boldsymbol{\nu})\,$ are  unitary non-equivalent for different values of  
$\boldsymbol{\nu}$. Precise definition and meaning of the parameters $\boldsymbol{\nu}\,$ which labels members of
the SW family will be given in the following sections. Here we only emphasize that the structure of the family, as well as the functional dependence of the Wigner functions on the coordinates of the symplectic manifold $\Omega_N\,,$ is
encoded in the type of degeneracy of the Stratonovich-Weyl operator kernel $\Delta(\Omega_N\,|\,\boldsymbol{\nu})\,.$ For example, if $\pi_i $ is an eigenvalue of the Hermitian  
$N\times N $ kernel $\Delta(\Omega_N)\,$ with the algebraic multiplicity $k(\pi_i)$, 
then its isotropy group $H$ is 
\begin{equation}
\nonumber H={U(k(\pi_1))\times U(k(\pi_2)) \times U(k(\pi_{r+1}))}\,,
\end{equation}
and the family of WF can be defined over the complex flag manifold
\begin{equation}\label{eq:Flag}
\Omega_N =\mathbb{F}^N_{d_1,d_2, \dots, d_r}=U(N)/H\,,
\end{equation}
where $(d_1, d_2, \dots, d_r)$ is a sequence of positive integers with sum
$N $, such that  $k(\pi_1)=d_1$ and $k(\pi_{i+1})=d_{i+1}-d_i$ 
with $d_{r+1}=N\,.$
In this case, the family of the Wigner functions of an 
$N$\--dimensional system in state $\varrho$ is constructed according to the Weyl rule:   \begin{eqnarray}
\label{eq:WFNK}
 W^{(\boldsymbol{\nu})}_{\varrho}
 (\boldsymbol{\vartheta}) &=&
 \mbox{tr}\left[ \varrho\, \Delta(\Omega_N\,|\,\boldsymbol{\nu})\right]\\
 &=&
 \text{tr}\left[\varrho\, X(\boldsymbol{\vartheta})
P^{(N)}(\boldsymbol{\nu}) X(\boldsymbol{\vartheta})^\dagger \right]\,,
\end{eqnarray}
where the classical counterpart  of density matrix $\varrho$ is given  by an $N\times N $ matrix $X(\boldsymbol{\vartheta})$ from the $d_{\mathbb{F}}$\--dimensional coset $\Omega_N$ with coordinates 
$\boldsymbol{\vartheta}=(\vartheta_1, \vartheta_2, \dots ,  \vartheta_{d_{\mathbb{F}}})$. 
The symbol $P^{(N)}(\boldsymbol{\nu})$ in Eq.~(\ref{eq:WFNK}) denotes a real diagonal $N\times N $ matrix, the entries of which are eigenvalues of the Hermitian kernel $\Delta(\Omega_N\,|\,\boldsymbol{\nu})$.

Our article is  organized as follows. In Section~\ref{sec-1}, based on the Stratonovich-Weyl correspondence,   ``master equations'' for the SW kernel matrix 
$\Delta(\Omega_{N}\,|\,\boldsymbol{\nu})$ will be derived and an ambiguity in the solution to these equations will be analyzed.
In Section~\ref{sec:IIIReduction}  connections between the proposed  generic SW mapping and a 
well-elaborated  $SU(2)$-symmetric 
spin-$j$ symbol correspondence (see e.g.  Ref.~\onlinecite{KlimovedeGuise2} and references therein) will be described.
It will be shown how to obtain the reduced Wigner function performing the reduction from flag manifold (\ref{eq:Flag}) to its 2-dimensional submanifold. 
Section~\ref{sec:IVExamples} is devoted to the exemplification of the suggested scheme of construction of the WF by considering two examples. We present a detailed description of the Wigner functions of 2 and 3-dimensional  systems, i.e., qubits and qutrits respectively. The reduced Wigner functions construction as well as the spin-1/2 and spin-1 Stratonovich-Weyl correspondence  derivation from the generic SW mapping will be done.  Our final comments and remarks  are given in Section \ref{sec:VConclusion}.

%%%%%%%%%%%%%%%%%%%%%%%%%%%%%%%%%
\section{The Wigner function via the Stratonovich-Weyl correspondence}
\label{sec-1}
%%%%%%%%%%%

\subsection{The Stratonovich-Weyl postulates}
\label{sec:IIA}

Let's consider an N-dimensional quantum system in a mixed state that is defined by the density matrix operator $\varrho$ acting on the Hilbert space $\mathbb{C}^N$. 
According to the basic principles of quantum mechanics, there is a mapping between the operators on the Hilbert space of a finite-dimensional quantum system and the functions on the phase space of its classical mechanical counterpart. This mapping can be realized with the aid of the Stratonovich-Weyl operator kernel $\Delta(\Omega_N)$
defined over a phase space $\Omega_N$. Particularly, the  Wigner quasiprobability distribution $W_\varrho(\Omega_N)$ corresponding to a density matrix $\varrho\,$  reads:
\begin{equation}\label{eq:WignerFunction}
W_\varrho(\Omega_N)= \mbox{tr}\left[\varrho \Delta(\Omega_N)\right]\,.
\end{equation}
The basic principles of quantum theory are accumulated in the following set of requirements  
(cf. formulation by Stratonovich~\cite{Stratonovich}, Brif and Mann~\cite{BrifMann1998Lett,BrifMann1999}) on SW kernel:
 \begin{enumerate}
\item[(I)] \underline{Reconstruction:} State $\varrho$ is reconstructed from the Wigner function  (\ref{eq:WignerFunction}) as
\begin{equation}\label{eq:DMWigner}
\varrho =\int_{\Omega_N} \mathrm{d}\Omega_N\, \Delta(\Omega_N) W_\varrho(\Omega_N) \,.
\end{equation}	
\item[(II)] \underline{Hermicity:} \hspace{0.5cm}{$\Delta(\Omega_N)= \Delta(\Omega_N)^\dagger $}.
\item[(III)] \underline{Finite Norm:} The state norm is given by the integral of the Wigner distribution 
\begin{equation}
\label{eq:FiniteNorm} 
\mbox{tr}[ \varrho ]= \int_{\Omega_N} \mathrm{d}\Omega_N W_\varrho(\Omega_N), 
\qquad
\int_{\Omega_N} \mathrm{d}\Omega_N\,\Delta(\Omega_N) = 1\,.
\end{equation}
\item[(IV)] \underline{Covariance:} The unitary transformations $\varrho^\prime = U(\alpha)\varrho U^\dagger(\alpha)$ induce the kernel change
\begin{equation}
\nonumber \Delta(\Omega_N^\prime) =U(\alpha)^\dagger\Delta(\Omega_N)U(\alpha).
\end{equation}
\end{enumerate}
For our further purposes it is worth to comment on measure in  (\ref{eq:DMWigner}). Identifying the phase space $\Omega_N$ as  a flag manifold
(\ref{eq:Flag}), the measure  in the reconstruction integral (\ref{eq:DMWigner}) 
can be written formally as 
\[
\mathrm{d}\Omega_N = C_N^{-1}{\mathrm{d}\mu_{SU(N)}}/{\mathrm{d}\mu_H}\,,
\]
where $C_N$ is a real normalization constant, 
$\mathrm{d}\mu_{SU(N)}$ 
%and 
%$\mathrm{d}\mu_{H}$ are
is the normalized Haar measures on the $SU(N)$. 
%group and the isotropy group $H$ respectively. 
Since the integrand in (\ref{eq:DMWigner}) is a function of
the coset variables only, the reconstruction integral can be extended to the whole group  $SU(N)$, 
\begin{equation}
\label{eq:reconstovergroup}
\varrho = Z_N^{-1}\int_{SU(N)} \mathrm{d}\mu_{SU(N)}\, \Delta(\Omega_N) W_\varrho(\Omega_N) \,,
\end{equation}	
by introducing the normalization constant $Z_N^{-1}= C_N^{-1}/\mbox{vol}(H)\,.$
Here, the factor $\mbox{vol}(H)$ denotes the volume of the isotropy group $H$ calculated with the measure induced by a given embedding, $H \subset SU(N)$.

Summarising all these commonly accepted  views, in the subsequent studies the following definition of SW kernel will be used.

\textbf{Definition 1.}
\textit{The  kernel satisfying  postulates (I)-(IV) and  providing  the  mapping from an element  of the space state $\varrho$ to the Wigner function (\ref{eq:WignerFunction}) is called the Stratonovich-Weyl kernel.}

%%%%%
\subsection{Master equations for Stratonovich-Weyl kernel }
\label{subsec-2.1}

Now it will be shown that the above generic  requirements on SW kernel can be reformulated in terms of simple algebraic equations. Namely, the following proposition takes place.  

\textbf{Proposition 1.}
\textit{The Stratonovich-Weyl kernel  $\Delta(\Omega_N)$
with isotropy group $H \in SU(N)\,,$ defined 
on a phase-space $\Omega_N$ identified as  a flag manifold $U(N)/H\,,$  satisfies the following algebraic equations}:
\begin{equation}\label{eq:ME}
\mbox{tr}\left[\Delta(\Omega_N)\right]=1\,,  \qquad  
\mbox{tr}[\Delta(\Omega_N)^2] = N\,.
\end{equation}

To prove the  {Proposition 1.}\,, note that relations  (\ref{eq:WignerFunction}) and  (\ref{eq:reconstovergroup}) imply the integral identity
\begin{equation}\label{eq:RhoIdentity}
\varrho=Z_N^{-1} \int_{SU(N)} \mathrm{d}\mu_{SU(N)}\,  \Delta(\Omega_N)\,
\mbox{tr}\left[\varrho\Delta(\Omega_N)\right]\,.
\end{equation}
To  proceed further we use the singular value decomposition of the Hermitian kernel $\Delta(\Omega_N)$: 
\begin{equation}\label{eq:diagkernel}
\Delta(\Omega_N)=U(\vartheta)PU^{\dagger}(\vartheta), \; P=\mbox{diag}||\pi_1,\pi_2,\dots,\pi_N||\,,
\end{equation}
with the following descending order of the eigenvalues 
\begin{equation}\label{eq:desorder}
\pi_1 \geq \pi_2 \geq \dots \geq \pi_N\,.  
\end{equation}
The unitary matrix $U(\vartheta)$ in (\ref{eq:diagkernel}) is not unique and the character of its arbitrariness follows from the  degeneracy  of the  spectrum $\sigma(\Delta)\,$ of the SW kernel,  i.e., by the isotropy group  $H\subset SU(N)$ of the diagonal  matrix  $P$.  
Thus, we assume that the diagonalizing matrix $U(\vartheta)$ belongs to a certain  coset  $U(N)/H\,.$ It is convenient  to identify it 
with a complex flag manifold (\ref{eq:Flag}) and use the coordinates $\vartheta_1, \vartheta_2, \dots ,  \vartheta_{d_{\mathbb{F}}}$ for its description .
 
Substituting $\Delta(\Omega_N) $ in  (\ref{eq:RhoIdentity})  with the decomposition (\ref{eq:diagkernel}), we get the identity, 
\begin{equation}\label{eq:int1temp}
Z_N^{-1}\int_{SU(N)}\mathrm{d}\mu_{SU(N)}(UPU^{\dagger})_{ik}(UPU^{\dagger})_{js}\varrho_{sj}=\varrho_{ik}.
\end{equation}
Now, performing the integration in identity (\ref{eq:int1temp}), we will get an algebraic equation for the SW kernel. Indeed, using the 4-th order Weingarten formula
\cite{Weingarten,Colins2003,ColinsSniady2006}:
\begin{widetext}
\begin{eqnarray}\label{IntUUdg}
\nonumber \int_{SU(N)} d\mu_{{}_{SU(N)}} U_{i_1 j_1} U_{i_2 j_2} U^{\dagger}_{k_1 l_1} U^{\dagger}_{k_2 l_2} =\frac{1}{N^2-1} \left(\delta_{i_1 l_1} \delta_{i_2 l_2} \delta_{j_1 k_1} \delta_{j_2 k_2}+\delta_{i_1 l_2}\delta_{i_2 l_1} \delta_{j_1 k_2} \delta_{j_2 k_1}\right)
\\ 
-\frac{1}{N(N^2-1)} \left( \delta_{i_1 l_1} \delta_{i_2 l_2} \delta_{j_1 k_2} \delta_{j_2 k_1} +  \delta_{i_1 l_2} \delta_{i_2 l_1} \delta_{j_1 k_1} \delta_{j_2 k_2}\right)\,,
\nonumber
\end{eqnarray}
\end{widetext}
on the left side of (\ref{eq:int1temp})  we arrive at the 
equations for the kernel:
\begin{eqnarray}\label{eq:cond1}
\left(\mbox{tr}[P]\right)^2&=&Z_N N\,,\\
\mbox{tr}[P^2]&=&Z_N N^2\,,
\label{eq:cond2}
\end{eqnarray}
Now  taking into account the finite norm condition (III) and the second order Weingarten formula
\begin{equation}
\nonumber \int_{SU(N)} \mathrm{d}\mu_{SU(N)}\, U_{i_1 j_1}  U^{\dagger}_{k_1 l_1}  = \frac{1}{N} \delta_{i_1 l_1} \delta_{j_1 k_1}\,, 
\end{equation}
one can verify that (\ref{eq:FiniteNorm}) is satisfied iff 
\begin{equation}\label{eq:standart}
\mbox{tr}[P]=Z_N N\,.
\end{equation}
Comparing (\ref{eq:standart}) with (\ref{eq:cond1}) allows to determine the normalization constant,  $Z_N=1/N$.
Finally, using the covariance condition (IV) and  $U(N)$ invariance of (\ref{eq:cond1})-(\ref{eq:cond2}), we  obtain the ``master equations'' for  the SW kernel:
\begin{equation}\label{eq:condNew1}
\mbox{tr}\left[\Delta(\Omega_N)\right]=1\,,  \qquad  
\mbox{tr}[\Delta(\Omega_N)^2] = N\,.
\end{equation}

%_______________________________

%\iffalse

\subsection{Dual picture}

Thus we come to the following dual description of finite-dimensional system with two basic ingredients, the quantum state space, the space of operators  $\mathfrak{P}_N$ on Hilbert space, 
and space
of matrix-valued functions $\mathfrak{P}^\ast_N$ on phase-space $\Omega_N\,.$ 
\textbf{Definition 2.} The quantum state space of $N\--$ dimensional system $\mathfrak{P}_N$ is the  following subspace of $N\times N$ matrices over $\mathbb{C}$: 
\begin{equation}\label{eq:StateSpace}
    \mathfrak{P}_N=\{ X \in M_N(\mathbb{C}) \ |\ X=X^\dagger\,,\quad  X \geq 0\,,  \quad \mbox{tr}\left( X \right) = 1   \}\,.
\end{equation}
\textbf{Definition 3.} The space $\mathfrak{P}^\ast_N$ of matrix-valued functions  on  phase-space $\Omega_N\,$ of $N\--$ dimensional system,
the so-called Stratonovich-Weyl kernel, is defined as:
\begin{equation}
\label{eq:SWspace}
    \mathfrak{P}^\ast_N=\{ X \in M_N(\mathbb{C}) \ |\ X=X^\dagger\,,\quad \mbox{tr}\left( X \right) = 1\,, 
    \quad
   \mbox{tr}\left( X^2 \right) = N 
    \}\,.
\end{equation}
\textbf{Definition 4.}  The Weyl dual pairing:
\begin{equation}
\label{eq:WignerFunction}
W_\varrho(\Omega_N) = \mbox{tr}\left[\varrho
\,\Delta(\Omega_N)\right]\,,
\end{equation} 
defines the Wigner quasiprobability function $W_\varrho(\Omega_N)$ on phase space $\Omega_N$ and  inverse mapping $\mathfrak{P}^\ast_N \to \mathfrak{P}_N$:
\begin{equation}\label{eq:Inverse Weyl}
\varrho =\int_{\Omega_N} \mathrm{d}\Omega_N\, \Delta(\Omega_N) W_\varrho(\Omega_N) 
\end{equation}	
for all  elements $\varrho \in \mathfrak{P}^\ast_N$ and $\Delta \in \mathfrak{P}^\ast_N\,.$

For further studies of this dual picture more detailed knowledge  of the structure of space $\mathfrak{P}^\ast$ is helpful.

%%%%%%%%%%%%%%%% SECTION II D%%%%%%%%%
\subsection{Space of solutions to the master equations}
\label{sec:IID}
%%%%%%%%%%%%%%%%%%%%%%%%%%%%

The following Proposition expounds a   structure of family of Wigner functions constructed from solutions to  (\ref{eq:condNew1}).

\textbf{Proposition 2.} \textit{ 
The  moduli space $\mathcal{P}_N$ of solutions to the master equations (\ref{eq:DMWigner}) represents  a spherical polyhedron   on   
$(N-2)\--$ dimensional sphere  $\mathbb{S}_{N-2}(1)\,$ of radius one.
The moduli space  $\mathcal{P}_N$ 
can be described  algebraically as follows. Let $O_{\boldsymbol{x}}$ be coadjoin orbit of $SU(N)$ parametererized by decreasingly ordered $n$-tuple $\boldsymbol{x}=(x_1, x_2,\dots, x_N)$ with components  summed up to zero, $\sum_{i}^N x_i=0$, 
and $C$ is positive Weyl chamber 
\begin{equation}
    C\colon = \{\, \boldsymbol{x} \in \mathbb{R}^N\,\ |\ \sum_{i}^N x_i=0\,, \ x_1 \geq x_2 \geq , \dots, \geq x_N \, \}\,,
\end{equation}
%\[
%C: \quad \{\mu \in \mathfrak{h}^\ast ,
%\langle \mu, \alpha \rangle \geq 0 ,  %\quad  \alpha \in \Delta^+ \} 
%\]
then the intersection of  $(N-1)\--$dimensional sphere, $\sum_{i}^N x^2_i =2\,,$ 
with the Weyl chamber  $C$
gives the moduli space $\mathcal{P}_N:$}
\begin{equation}
\label{eq:ModuliSpace}
     \mathcal{P}_{N} \simeq 
     C \cap\, \mathbb{S}_{N-1}( \sqrt{2} )\,.
 \end{equation}
To get convinced in the above statement, note that equations  (\ref{eq:condNew1}) impose two conditions  on eigenvalues of SW kernel  only.  Therefore, assuming the existence of $N$ different eigenvalues of SW kernel,   
the maximal number of continuous parameters $\boldsymbol{\nu}=(\nu_1, \nu_2, \dots, \nu_{N-2})$ characterizing the solution $\Delta(\Omega_N\,|\, \boldsymbol{\nu})$ can be  $N-2\,.$ Furthermore in addition, consider the SVD decomposition for $\Delta(\Omega_N\,|\, \boldsymbol{\nu})$,  with its diagonal part expanded over the basis elements of a Cartan subalgebra $ \mathfrak{h} \in \mathfrak{su}(N)$ 
\begin{equation}
\label{eq:SWkernelexp}
 \Delta(\Omega_N|\boldsymbol{\nu})=\frac{1}{N}U(\Omega_N)\left[I+\kappa\sum_{\lambda\in \mathfrak{h} }\mu_s(\boldsymbol{\nu})\lambda_s\right]U(\Omega_N)^\dagger,
\end{equation}
where $\kappa=\sqrt{{N(N^2-1)}/{2}}$\,, and the orthonormal basis $\{\lambda_1, \lambda_2, \dots,\lambda_{N^2-1}\}$ of the algebra $\mathfrak{su}(N)$ with respect to trace form $\mbox{tr}(\lambda_i\lambda_j)=2\delta_{ij}$ is chosen. 
Substitution  of (\ref{eq:SWkernelexp})
to the equation (\ref{eq:condNew1}) leads to the constraint on a real coefficients $\mu_s(\boldsymbol{\nu})\,,$: 
 \begin{equation}
 \label{eq:moduliWF}
 \sum_{s=2}^{N}\mu^2_{s^{2}-1}(\boldsymbol{\nu}) = 1\,.  
 \end{equation}
Finally, taking into account an  expansion of the
eigenvalues $x_1, x_2, \dots, x_N$ of the traceless part of SW kernel 
over the coefficients $\mu_3, \mu_8, \dots, \mu_{N^2-1} $, 
the equation  (\ref{eq:moduliWF}) reduces to  $\sum_{i}^N x^2_i =2\,,$ proving representation (\ref{eq:ModuliSpace}) for moduli space $\mathcal{P}_N$.
Hence, when  SW kernel is a generic one,  i.e., its isotropy group  is $U(1)^N\,, $ the master equations (\ref{eq:condNew1}) admits an 
$(N-2)\--$parametric solutions for  $\mu(\boldsymbol{\nu})$ with a real parameters $\boldsymbol{\nu}\,,$ which can be  chosen as $N-2$ spherical angles. Now, in order to determine the fundamental domain/the moduli space $\mathcal{P}_N\,,$ as the locus of points on sphere $\mathbb{S}_{N-2}(1)$ which are  in one-to-one correspondence with a given order of eigenvalues of  $\Delta(\Omega_N\,|\,\boldsymbol{\nu})$\,, it is necessary to restrict the range of the spherical angles.  After  restriction to  a  
subset $\mathcal{P}({\boldsymbol{\nu}}) \subset \mathbb{S}_{N-2}(1)\,,$ an ambiguity  of ordering of  the eigenvalues in SVD decomposition (\ref{eq:SWkernelexp}) is eliminated.  
Geometrically, fixing certain ordering of   eigenvalues (\ref{eq:desorder})  results in cuting out  the moduli space of  $\Delta(\Omega_N\,|\,\boldsymbol{\nu})$ in the form of a spherical polyhedron on $ \mathbb{S}_{N-2}(1)\,$. 
\cite{footnote}
The faces, edges and vertices of this polyhedron correspond to the SW kernels the isotropy group of which is larger than the maximal torus. 
 
%%%%%%%%%%%%%%%%%%%%%%%%%%%%%%%%%%%%%%
\subsection{Parameterizing the Wigner function}

Summarizing, we are in position to  make the following assertion. 

\textbf{Proposition 3: }
\textit{Consider the symplectic manifold $\Omega_N \simeq U(N)/U(1)^N\,$ and suppose that a  quantum $N\--$level system is in a mixed state
$\varrho\,$
characterized by $N^2-1$\--dimensional Bloch vector
$\boldsymbol{\xi}\,,$
\begin{equation}
\label{eq:rhoN}
\varrho =\frac{1}{N}\left(I + \sqrt{\frac{N\left(N-1\right)}{2}}\left(\boldsymbol{\xi},\boldsymbol{\lambda}\right)\right)\,.
\end{equation}
The SW mapping implemented by SW kernel   $\Delta(\Omega_N|\boldsymbol{\nu})$ defines 
a family of the Wigner functions 
\begin{equation}\label{eq:WFCartan}
W^{(\boldsymbol{\nu})}_{\boldsymbol{\xi}} (\theta_1,\theta_2, \dots,  \theta_d)=\frac{1}{N}\left[1 + \frac{N^2-1}{\sqrt{N+1}}\,(\boldsymbol{n}, \boldsymbol{\xi})\right]\,,
\end{equation}
where $\boldsymbol{n}$ is $(N^2-1)$\--dimensional  unit vector
\begin{eqnarray}
\label{eq:n}
\boldsymbol{n} = \mu_3(\boldsymbol{\nu}) 
\boldsymbol{n}^{(3)} + \mu_8(\boldsymbol{\nu}) \boldsymbol{n}^{(8)}+\dots+ 
\mu_{N^{2}-1}(\boldsymbol{\nu})\boldsymbol{n}^{(N^{2}-1)}\,.  
\end{eqnarray}
The vector $\boldsymbol{n}$ in (\ref{eq:n}) 
is decomposed into $(N-1)$ orthogonal vectors $\boldsymbol{n}^{(3)}\,,  \boldsymbol{n}^{(8)}\,, 
\dots, \boldsymbol{n}^{(N^{2}-1)}\,.$ These  vectors  correspond to the basis elements $\lambda_{s^2-1} \in \mathfrak{h}\,, s=2,3,\dots,N\,$ of the Cartan subalgebra  $\mathfrak{h}$ and are given as:
\begin{equation}
\nonumber \boldsymbol{n}^{(s^2-1)}_\mu = \frac{1}{2}\,\mbox{tr}\left( U\lambda_{s^2-1}U^\dagger\lambda_\mu \right)\,, \quad
\lambda_{s^2-1} \in \mathfrak{h}\,.
\end{equation}
The real coefficients  $\mu_1(\boldsymbol{\nu}), \mu_2(\boldsymbol{\nu}), \dots, \mu_{N^2-1}(\boldsymbol{\nu})$
in (\ref{eq:n}) are coordinates of SW kernel on the moduli space $\mathcal{P}_N({\boldsymbol{\nu}})\,.$
}

As it was mentioned in the Introduction, the number $d(N)$ of independent variables $\vartheta $ in the Wigner function  (\ref{eq:WFCartan}) depends on the  isotropy group of SW kernel. 
The maximal number for a given $N$ equals to  $\max{d(N)} = N(N-1)\,$ and corresponds to  the maximal torus $T\,.$ 
However, depending on the symmetry of SW kernel and the state, the number of independent variable in WF can be reduced.  
We leave a complete analysis of construction of WF for system with symmetry group $g \subset U(N)$  for a future publication. However, in subsequent sections we will exemplify in detail some generic features of a reduction process considering the Wigner functions of the lowest dimensional systems, $N=2$ and $N=3$, a single qubit and a single qutrit.

\noindent{$\bullet $\,\bf  Comments on a set of conditions for SW kernel\,$\bullet $} Finalizing our derivation of master equations, it is worth commenting on the particular formulation of Stratonovich-Weyl correspondence rules which we use in this paper. 
In our notations, the Stratonovich properties (cf. also Refs.~\onlinecite{BrifMann1998Lett,BrifMann1999}) read:  
\begin{enumerate}
\item Linearity:  $A \to W_A(\Omega_N)$ is one-to-one map;
\item  Standardisation:
  \begin{equation}
\nonumber  Z_N^{-1}\int \mathrm{d}\mu_{SU(N)} W^{(\boldsymbol{\nu})}_A(\Omega_N)= \mbox{tr}[A]\,,
  \end{equation}
\item Covariance:  under transformation of operators  $A^g=g^\dagger A g\,,$ the symbol changes as  $W^{(\boldsymbol{\nu})}_{A^g}(\Omega_N)=W^{(\boldsymbol{\nu})}_{A}(g\cdot\Omega_N)\,,$ 

\item Traciality:
\begin{equation}\label{eq:traciality}
Z_N^{-1}\int \mathrm{d}\mu_{SU(N)} W^{(\boldsymbol{\nu})}_A(\Omega_N)W^{(\boldsymbol{\nu})}_B(\Omega_N) = \mbox{tr}[AB]\,.
\end{equation}
\end{enumerate}
Comparing this list  with
the requirements (I)-(IV) in Section \ref{sec:IIA}\,, one can see that in this note we prefer to fix directly the reconstruction  formula (I) as the inverse  Weyl rule with the same SW kernel $\Delta(\Omega)$ used in the construction of WF in (\ref{eq:WignerFunction}).
In other words, in the present article we are discussing the non-uniqueness of the Wigner quasiprobability, i.e., functions with ``self-dual'' SW kernel.
Here it is worth mentioning that the traciality condition (\ref{eq:traciality}) is satisfied automatically for SW symbols of operators constructed with the aid of ``self-dual'' kernel $\Delta(\Omega_N\,|\,\boldsymbol{\nu})$ for all $\boldsymbol{\nu}\,.$
Indeed, once again the usage of  the Weingarten formula for evaluation of 
the integral (\ref{eq:traciality}) results in identity modulo the  ``master equations''.
%\fi

%%%%%%%%%%%%%.  NEW SECTION %%%%%%%%%%%%%%%%%%%
\section{Reduction to  $SU(2)$ symmetric spin-$j$ correspondence}
\label{sec:IIIReduction}

This section emerged as a result of our reply to the Referee's suggestion to clarify connections between the proposed  generic SW mapping and a 
well-elaborated  $SU(2)$-symmetric 
spin-$j$ symbol correspondence. To make presentation  self-sufficient, we  start with the definitions of spin-$j$ system and a spin-$j$ symbol correspondence in the form borrowed from the book by de Rios and Straum  \cite{deRiosStraum}. 

\textbf{Definition 5.}
\textit{
A spin-$j$ system is a complex Hilbert space $H_j \simeq \mathbb{C}^{N}$ together with an irreducible unitary representation
$$
\phi_j\colon\  SU(2) \to  G \subset U(H_j) \simeq  U(N)\,, \qquad   N=2j+1 \in \mathbb{N}\,,
$$
where $G$ denotes the image of $SU(2)$ which is isomorphic to $SU(2)$ or $SO(3)$
according to whether $j$ is half-integral or integer.
}

\textbf{Definition 6.}
\textit{
A symbol correspondence for a spin-$j$ system is a rule which associates to each operator $P \in \mathcal{B}(\mathcal{H}_j)$ a smooth function $W^j_P$ on $\mathbb{S}^2$, called its symbol, with the following
properties:}
\begin{enumerate}
    \item[(i)] Linearity:  the map $P \to W^j_P$ is linear and injective;
    \item[(ii)] Equivariance: $W^j_{P^g}=\left(W^j_P \right)^g$, for each $g \in SO(3)$\,;
    \item[(iii)] Reality: $W^j_{P^\dagger}(\boldsymbol{n})=\overline{W^j_P(\boldsymbol{n})}$\,;
    \item[(iv)] Normalization:
$\frac{1}{4\pi}\,\int_{\mathbb{S}^2}\,W^j_P(\boldsymbol{n})\mathrm{d}S=\frac{1}{N}\mbox{tr}(P) $\,;
\end{enumerate}

\textbf{Definition 7.}
\textit{A Stratonovich-Weyl correspondence is a symbol correspondence
that additionally to (i)-(iv) axioms also satisfies the so-called isometry axiom:}
\begin{enumerate}
    \item[(v)] Isometry:
$\langle W^j_P W^i_Q\rangle =\frac{1}{N}\mbox{tr}\left(P^\dagger Q\right) $\,.
\end{enumerate}
\textit{The left-hand side of the equations denotes the normalized $L^2$ inner product of two functions on the sphere,
$$
\langle F_1, F_2\rangle = \frac{1}{4\pi}\,\int_{\mathbb{S}^2}\overline{F_1(\boldsymbol{n})}F_2(\boldsymbol{n})\mathrm{d}S\,.
$$
}

{\bf Proposition 4:\ }
\textit{
For any $N=(2j+1)$, where $j=\frac{1}{2}, 1, \frac{3}{2},  \dots ,$ among solutions 
to the ``master equations''
(\ref{eq:ME}) one can always 
find at least one SW  kernel $ \Delta^{(\boldsymbol{k})}$, of a symmetry type
$[H_{\boldsymbol{k}}]\,,$  such that a generic dual pairing  (\ref{eq:WignerFunction}) with a density matrix $\varrho_{(\boldsymbol{q})}$ of  $[H_{\boldsymbol{q}}]$
symmetry type reduces to the $SU(2)$-symmetric spin\--$j$ correspondence.
The  reduced Wigner function 
$W^{(\boldsymbol{k})}_{\ (\boldsymbol{q})}\,$ 
associated to a density matrix is 
defined either on a 1-dimensional subspace of phase-space for half-integer,  $j=
\frac{1}{2}\,, \frac{3}{2}\,,  \dots , $ 
or on 
a 2-dimensional subspace of phase-space  for integer, $j=1\,, 2\,,  \dots $
}

{\bf Proposition 5:\ }
\textit{The  reduced Wigner quasiprobabily distribution $W^{(\boldsymbol{k})}_{\ (\boldsymbol{q})}$, when the symmetry groups of density matrix and SW kernel correspondingly are $H_{\boldsymbol{k}}$  and $H_{\boldsymbol{q}}$\, ,   can be determined as follows.}  
\begin{itemize}
    \item \textit{
Introduce the double coset, 
 $  \mathbb{B}^{N}_{\boldsymbol{k}, \boldsymbol{q}} = H_{\boldsymbol{q}} \backslash  SU(N) / H_{\boldsymbol{k}}\,$
with the following left and right factors:
\begin{eqnarray}
\label{eq:Kk}
 H_{\boldsymbol{k}}&=&
S\left(
U(k_1)\times U(k_2)\times \cdots \times U(k_{L})
\right)\,, \qquad \prod_{i=1}^{L} \det\left(U(k_i)\right)=1\,, \quad  \sum_{i=1}^{L}k_i=N\,,\\
H_{\boldsymbol{q}}&=&
S\left(
U(q_1)\times U(q_2)\times \cdots \times U(q_{R})
\right)\,, \qquad \prod_{i=1}^{R} \det\left(U(q_i)\right)=1\,, \quad 
\sum_{i=1}^{R}q_i = N\,,
\label{eq:Kq}
\end{eqnarray}
where  ${\boldsymbol{k}}=(k_1, k_2, , \dots k_L)$ and 
${\boldsymbol{q}}=(q_1, q_2, , \dots q_R)\,$
are degrees of degeneracy of  the decreasingly ordered eigenvalues  of a given density matrix and SW kernel,
$r_1> r_2 > \dots  > r_L\,$ and $
    \pi_1 > \pi_2 > \dots  > \pi_R\,,
$
\begin{eqnarray}
\label{eq:rhospec}
\boldsymbol{r}^{\downarrow}&=&\mbox{spec}\{
\overbrace{(r_1, \dots, r_1)}^{k_1}\,;\, \overbrace{(r_2, \dots, r_2)}^{k_2}\,;\, \dots \,;\, \overbrace{(r_L, \dots, r_L)}^{k_L}\}\,,
\\
\pi^{\downarrow}&=&\mbox{spec}\{ \overbrace{(\pi_1, \dots, \pi_1)}^{q_1}\,;\, 
\overbrace{(\pi_2, \dots, \pi_2)}^{q_2}\,;\, \dots \,;\, \overbrace{(\pi_R, \dots, \pi_R)}^{q_R}\}\,,
\label{eq:deltaspec}
\end{eqnarray}
\item Consider a mapping from 
 $  \mathbb{B}^{N}_{\boldsymbol{k}, \boldsymbol{q}}\,$
to the subspace of the Birkhoff polytope $B_N\,,$ by prescribing to each element $Z \in  \mathbb{B}^{N}_{\boldsymbol{k}, \boldsymbol{q}}$ the  unistochastic matrix: \begin{equation}
\label{eq:Birhoff}
 \mathbb{B}^{N}_{\boldsymbol{k}, \boldsymbol{q}} \to B_N\colon  \qquad   B_{ij}=|Z_{ij}|^2\,, \qquad
 \forall Z \in \mathbb{B}^{N}_{\boldsymbol{k}, \boldsymbol{q}}\,,
\end{equation}
\item Define, based on the above mapping (\ref{eq:Birhoff}),  the bilinear form: 
\begin{equation}
\label{eq:WFB1}
W^{(\boldsymbol{k})}_{\ (\boldsymbol{q})} = r^\downarrow_i B_{ij}\pi^\downarrow_j =\left(\boldsymbol{r}^{\downarrow}, \boldsymbol{\pi}^{\downarrow}\right)_{B}\,.
\end{equation} 
}
\item \textit{Find pairs of tuples 
$\boldsymbol{k}^\downarrow =(k_1, k_2, , \dots k_L)$ and 
$\boldsymbol{q}^\downarrow =(q_1, q_2, , \dots q_R)\,$ that solve the equation (\ref{eq:Even1})  if $j$ is a half-integer  or (\ref{eq:1Odd}) if $j$ is an integer.}
\end{itemize}
\textit{
As a result, the pairing  (\ref{eq:WFB1}) for each pair of solution  $(\boldsymbol{k}^\downarrow\,,  \boldsymbol{q}^\downarrow)$ defines the reduced WF,  which realizes $SU(2)$ symmetric spin\--$j$ correspondence. The variety of possible symmetries of SW correspondence is  determined by all pairs of Young diagrams corresponding to a set of solutions $\{\boldsymbol{k}^\downarrow\,,  \boldsymbol{q}^\downarrow\}$.
(The symmetry of a point $x \in \mathfrak{P}^{\ast}$ 
associated with the adjoint action of group  $G$
is given by the isotropy (stability)  group $G_x$:
\[
G_x = \{ g \in G\, | \, x = g^{-1} 
x \, g\,  
\}\, ).
\]
The reduced WF corresponding to another ordering of eigenvalues  $\boldsymbol{r}\,$ obtained by transposition $P$ from $\boldsymbol{r}^\downarrow$  is given by pairing (\ref{eq:WFB1}) with  transposed matrix: 
\begin{equation}
\label{eq:transB}
    {B}_P = P B\,.
\end{equation}
The result of transposition (\ref{eq:transB}) can be dragged to the change of a phase space coordinates.  
}

To prove the {\bf Propositions}, consider SVD decomposition for density matrices $\varrho_{({\boldsymbol{q}})} $ and SW kernel $\Delta^{(\boldsymbol{k})}$  with spectrum of the  types of degeneracy (\ref{eq:rhospec}) and (\ref{eq:deltaspec}): 
\begin{equation}
\label{eq:SVDAB}
\varrho_{({\boldsymbol{q}})}=V 
\begin{pmatrix}
r_1\mathbb{I}_{k_1} & \cdots & 0 \\
\vdots& \ddots & \vdots \\
0 & \cdots & r_L\mathbb{I}_{k_L}
\end{pmatrix} V^\dagger\,, \qquad 
\Delta^{(\boldsymbol{k})}= U \begin{pmatrix}
\pi_1\mathbb{I}_{q_1} & \cdots & 0 \\
\vdots& \ddots & \vdots \\
0 & \cdots & \pi_R\mathbb{I}_{q_R}
\end{pmatrix} U^\dagger\,.
\end{equation}
These are  not unique. The most general family of a diagonalizing unitary matrices $V$ and $U$ in (\ref{eq:SVDAB}) read 
\begin{equation}
\label{eq:GenericVW}
V = V^{\downarrow}
\begin{pmatrix}
V_{k_1} & \cdots & 0 \\
\vdots& \ddots & \vdots \\
0 & \cdots & V_{k_L}
\end{pmatrix}
P\,,
\qquad
U = U^{\downarrow}
\begin{pmatrix}
U_{q_1} & \cdots & 0 \\
\vdots& \ddots & \vdots \\
0 & \cdots & U_{q_R}
\end{pmatrix}
Q\,,
\end{equation}
where  $V^{\downarrow}$ and 
 $ U^{\downarrow}$ denote the unitary matrices constructed of a right eigenvectors 
of matrix $\varrho $ and $\Delta$\, ordered according to their decreasing eigenvalues.  
The matrices $V_{k_1},  \dots , V_{k_L} $
and 
$U_{q_1},  \dots , U_{q_R} $
are arbitrary unitary matrices of  order $k_1,  \dots, k_L$ and 
$q_1,  \dots, q_R$
respectively, 
$P$ and $Q$ are matrices 
transposing  the columns.

Let us make a few statements on  an unistochasttic matrices 
in (\ref{eq:Birhoff}). 
Note that matrices $B_{ij}$ form a subset of space $\mathcal{U}_N$ of the so-called  \textit{unistochastic matrices}. Its  dimension reads  
\begin{equation}
\label{eq:dimUNI}
 \dim\mathcal{U}_N=(N-1)^2\,.   
\end{equation}
Now, first of all, we are ready to show that WF  for a most generic SW kernel and density matrices has $(N-1)^2$ dimensional support in accordance with dimension of the space of unistochastic matrices, 
(\ref{eq:dimUNI}). 
Indeed, 
taking into account that for a generic case, without symmetries, the isotropy groups of states   and SW kernel are minimal ones, 
\[
\dim H_{\boldsymbol{q}}
 =\dim H_{\boldsymbol{k}}=N-1\,,
\]
a real dimension of the coset $\mathbb{B}_{\boldsymbol{k},\boldsymbol{q}}$: 
\begin{eqnarray}
 \dim\mathbb{B}^N_{\boldsymbol{k}, \boldsymbol{q}} &=& N^2 -1 -\dim H_{\boldsymbol{q}}
 -\dim H_{\boldsymbol{k}}\,  
\end{eqnarray}
reduces for a generic case to  
\begin{eqnarray}
 \dim\mathbb{B}^N_{\boldsymbol{k}, \boldsymbol{q}}\bigl|_{\sf Generic} &=& N^2 -1 -2(N-1)=(N-1)^2\,.  
\end{eqnarray}

A realization of the $SU(2)$-symmetric SW correspondence for spin-j assumes that  $N=2j+1$ level system 
is in specific states possessing  a nontrivial isotropy group $H_{\boldsymbol{q}}$ and at the same time  SW kernel   has a symmetry given by a certain isotropy group $H_{\boldsymbol{k}}$
as well.

Now, in order to determine both symmetry groups, we formulate the set of algebraic equations  for $\boldsymbol{k}$ and $\boldsymbol{q}$ tuples. It turns out that these  equations are different for systems with odd and even number of levels.
Namely, the equations for a half-integer spins,  $j=
\frac{1}{2}\,, \frac{3}{2}\,,  \dots ,
$
are  
\begin{equation}
\label{eq:DC1sphere}
    \dim\mathbb{B}^{2j+1}_{\boldsymbol{k}, \boldsymbol{q}}=1\,,
\end{equation}
while  for an integer spins,  $j=1\,, 2\,,  \dots, $ read  
\begin{equation}
\label{eq:DC2sphere}
    \dim\mathbb{B}^{2j+1}_{\boldsymbol{k}, \boldsymbol{q}}=2 \,.
\end{equation}
Using the expression for the coset dimension: 
\begin{eqnarray}
 \dim\mathbb{B}_{\boldsymbol{k}, \boldsymbol{q}} &=& N^2 -1 -\dim H_{\boldsymbol{q}}
 -\dim H_{\boldsymbol{k}}\\&=&
 4j(j+1)-\sum_{i=1}^{L}k_i^2 -\sum_{i=1}^{R}q_i^2+2\,,  
\end{eqnarray}
we reformulate (\ref{eq:DC1sphere}) and (\ref{eq:DC2sphere}) as the problem of solving the equations: 
\begin{enumerate}
    \item For a half-integer ($N$-even), 
 \begin{eqnarray}
 \label{eq:Even1}
    &&\sum_{i=1}^{L}k_i^2 +\sum_{i=1}^{R}q_i^2 =1+4j(j+1)\,,
    \\
    &&
    \sum_{i=1}^{L}k_i=\sum_{i=1}^{R}q_i =
    2j+1\,.
    \label{eq:Even2}
\end{eqnarray}
\item For an integer spin ($N$-odd), 
 \begin{eqnarray}
 \label{eq:1Odd}
    &&\sum_{i=1}^{L}k_i^2 +\sum_{i=1}^{R}q_i^2=4j(j+1)\,,
\\
    &&
    \sum_{i=1}^{L}k_i=\sum_{i=1}^{R}q_i =
    2j+1\,,
    \label{eq:2Odd}
\end{eqnarray}
\end{enumerate}
with respect to  natural numbers $(\boldsymbol{k}, \boldsymbol{q})\,.$  
Hence, a proof of the 
{\bf Proposition} reduces to  
determination of solutions of the above algebraic equations.
 We do not have a complete solution to this problem for an arbitrary $N$\,, but  
it is  straightforward to find 
a set of solutions for  low values of spin-$j$. The results for  $1 \leq j \leq 7/2$  are given in Table \ref{Table1} and 
Table \ref{Table2}.

%%%%%%%%%%%%%%%%%%%%%%%%% TABLE for half-integer spins 
\begin{table}[h!]
\begin{center}
\begin{tabular}{||c| c| c |c|c|} 
\hline
%\rowcolor{yellow}
\multicolumn{5}{|c|}{} \\ 
%\rowcolor{yellow}
\multicolumn{5}{|c|}{\bf \large List of solutions for a half-integer spin } \\[3.1ex]  
 \hline\hline
 SPIN & SW KERNEL DEGENERACY & STATE DEGENERACY  & P(N)& $4j(j+1)$  \\
  j &$(k_1, k_2,  \cdots ,  k_{L-1}, k_L)$ &
 $ (q_1,  q_2, \cdots ,  q_{R-1}, q_R)$ &&\\
 \hline\hline
 1/2   & (1,1) & (1,1) & 2 & 3 \\
\hline 
 3/2 & (3,1,)& (2,1,1)  & 5 & 15 \\
\hline
\hline
%  5/2
\multicolumn{1}{||c|}{\multirow{4}{*}{5/2}}&
(4,1,1) & (4,1,1) & \multicolumn{1}{|c|}{\multirow{4}{*}{11}} &\multicolumn{1}{|c|}{\multirow{4}{*}{35}} \\
\cline{2-3}
& (3,3) & (3,3) &  &  \\
\cline{2-3}
& (4,1,1) & (3,3) &  &  \\
\cline{2-3}
\cline{2-3}
& (5,1) & (2,2,1,1) &  &  \\
\hline
\hline
% 7/2
\multicolumn{1}{||c|}{\multirow{5}{*}{7/2}}&
(7,1) & (3,1,1,1,1,1)  &\multicolumn{1}{|c|}{\multirow{5}{*}{22}}  &\multicolumn{1}{|c|}{\multirow{5}{*}{63}}\\
\cline{2-3}
 & (7,1)  &(2,2,2,1,1)  &  &  \\ 
 \cline{2-3}
 & (6,2) & (4,2,2)  &  &  \\
 \cline{2-3}
 & (6,1,1) & (4,3,1) &  &  \\
 \cline{2-3}
 & (5,3) & (5,2,1)  &  &  \\
  \cline{2-3}
 & (4,4) & (4,4)  &  &  \\
\hline
\end{tabular}
\end{center}
\caption{Symmetries and partitions corresponding to low dimensional half-integer $SU(2)$-symmetric  spin-j correspondence.}
    \label{Table1}
\end{table}
%%%%%%%%%%%%%%%%%% END of TABLE

In the tables $P(N)$ is a partition function  which gives a number of possible partitions of a non-negative integer  ${N}$ into natural numbers.

%%%%%%%%%%%%%%%%%%%%%%%%% TABLE For Integer spin 
\begin{table}[h!]
\begin{center}
\begin{tabular}{||c| c| c |c|c|} 
\hline
%\rowcolor{yellow}
\multicolumn{5}{|c|}{} \\ 
%\rowcolor{yellow}
\multicolumn{5}{|c|}{\bf \large List of solutions for an integer   spins } \\[3.1ex]  
 \hline\hline
 SPIN & SW KERNEL DEGENERACY & STATE DEGENERACY  & P(N)& $4j(j+1)$  \\
  j &$(k_1, k_2,  \cdots ,  k_{L-1}, k_L)$ &
 $ (q_1,  q_2, \cdots ,  q_{R-1}, q_R)$ &&\\
 \hline\hline
 1   & (2,1) & (1,1,1) & 3& 8 \\
% 2
\hline
\hline
\multicolumn{1}{||c|}{\multirow{2}{*}{2}}&
(4,1) & (2,1,1,1) & \multicolumn{1}{|c|}{\multirow{2}{*}{7}} &\multicolumn{1}{|c|}{\multirow{2}{*}{24}} \\
 \cline{2-3}
 & (3,1,1)& (3,2) & &  \\
 \hline
\hline
% 3
\multicolumn{1}{||c|}{\multirow{4}{*}{3}}&
 (6,1)& (2,2,1,1,1) & \multicolumn{1}{|c|}{\multirow{4}{*}{15}} & \multicolumn{1}{|c|}{\multirow{4}{*}{48}}\\
\cline{2-3}
& (5,2) &(3,3,1)  &  &  \\  
\cline{2-3}
&(5,2) &(4,1,1,1)  &  &  \\ 
\cline{2-3}
&(5,1,1)  & (4,2,1) &  &  \\ 
\hline
\hline
\end{tabular}
\end{center}
\caption{Symmetries and partitions corresponding to low dimensional integer $SU(2)$-symmetric  spin-j correspondence.}
\label{Table2}
\end{table}
%%%%%%%%%%%%%%%%%% END of TABLE

%%%%%%%%%%%%%%%%%%.  EXAMPLES %%%
\section{Examples}
\label{sec:IVExamples}

In this section we consider in detail examples of low-dimensional quantum systems. The explicit form of Wigner functions  for $N=2$ and $N=3$ level system  will be  given. Apart from these,  we  will describe the reduction of the Wigner functions  to the   subspaces of phase space  constructing the  SW mapping when systems possess a certain symmetry. 
The construction  of the reduced WF of spin-1/2  and spin-1  is presented.

%%%%%%%%%%%%%%%%%%%%%%%%%%%%%%%%%% QUBIT
\subsection{Wigner function of a single qubit}
%----------------------------------------------

\noindent{$\bullet $\,\bf  A qubit mixed state \,$\bullet $} Consider a generic 2-level system in a mixed state, characterized by the Bloch vector $\boldsymbol{r}$  with spherical components, 
$
\boldsymbol{r}=
r(\sin{\alpha^\ast}\cos{\beta^\ast}\,,  \sin{\alpha^\ast}\sin{\beta^\ast}\,,
\cos{\alpha^\ast})\,,
$ 
\begin{equation}
\varrho =
\frac{1}{2}\mathbb{I}+\frac{1}{2}\,(\boldsymbol{r},\boldsymbol{\sigma})\,,
\end{equation}
where the vector $\boldsymbol{\sigma}$ refers to the set of the Pauli matrices,  $ \boldsymbol{\sigma}=(\sigma_1, \sigma_2, \sigma_3)\,.$ 
 Equivalently, $\varrho$ in SVD form reads: 
\begin{equation}
    \varrho = V(\alpha^\ast, \beta^\ast)\left(
\begin{array}{cc}
 r_1 & 0  \\
 0 & r_2
\end{array}
    \right)V(\alpha^\ast, \beta^\ast)^\dagger\,.
\end{equation}
The eigenvalues of the density matrix $r_1$ and $r_2$ are linear combinations of the radius $r$ of the Bloch vector: 
\[
r_1=\frac{1}{2}(1+r)\,,\qquad
r_2=\frac{1}{2}(1-r)\,,\qquad
\]
and matrix $V$ is an element of the coset 
$SU(2)/U(1)$ in conventional parameterization, 
\begin{equation}
    V(\alpha^\ast, \beta^\ast)=
    \exp\left(i\frac{\alpha^\ast}{2}{\sigma}_3\right)\exp\left(i\frac{\beta^\ast}{2}{\sigma}_2\right)
    \exp\left(-i\frac{\alpha^\ast}{2}{\sigma}_3\right)\,.
\end{equation}

%%%%%%%%%%%%%%%% QUBIT SW KERNEL 
\noindent{$\bullet $\,\bf SW kernel\,$\bullet$} The master equations (\ref{eq:ME}) give a unique solution for the spectrum of 2-dimensional SW kernel:
\begin{equation}
\mbox{spec}\left(\Delta(\Omega_2)\right)=\Big|\Big|\,\frac{1+\sqrt{3}}{2}\,,\,\frac{1-\sqrt{3}}{2}\,\Big|\Big|\,.
\end{equation}
Therefore,  SW kernel  of qubit 
\begin{equation}
\label{eq:SW2}
\Delta(\Omega_2) =\frac{1}{2}\,U(\Omega_2)\left(
\begin{array}{cc}
 1+\sqrt{3} & 0  \\
 0 & 1-\sqrt{3}
\end{array}
    \right)U^\dagger(\Omega_2)=\frac{1}{2}\,
    \mathbb{I} +\frac{\sqrt{3}}{2}\,(\boldsymbol{n},\boldsymbol{\sigma})\,,
\end{equation}
is defined over 2-sphere described by the unit vector,
$
{n}_i=U(\Omega_2)\,\sigma_3 U(\Omega_2)^\dagger = (\sin\theta\cos\phi,  \sin\theta\sin\phi, \cos\theta)\,.$
Hence, the Wigner function for 2-level system in a state $\varrho $ on a 2-sphere reads: 
 \begin{equation}
     W_\varrho(\boldsymbol{n})=\frac{1}{2} + \frac{\sqrt{3}}{2}\,
     (\boldsymbol{r},\boldsymbol{n})\,, \qquad 
     \boldsymbol{n} \in \mathbb{S}^2\,.
 \end{equation}

%$$\boldsymbol{n}=\left(\cos (\pi -\alpha ) \sin \beta ,\sin (\pi %-\alpha ) \sin \beta ,\cos \beta \right)$$.

%%%%%%%%%%%%% WF 
\noindent{$\bullet $\,\bf Reduced  WF of qubit\,$\bullet$}
For the case of qubit the symmetry analysis is trivial. Two-level system is  associated with spin-1/2 system directly.   For spin-1/2 there are only  $P(2)= 2$ partitions, namely $(1,1)$ and $(2)$.
%presented by the following %Young diagrams: \\

%$j=\frac{1}{2}$ 
%\quad
%\quad \yng(1,1)
%\quad \yng(2) 

According to  (\ref{eq:Even1}), the partition (1,1) gives sought symmetric coset with a same left and right factors $S(U(1)\times U(1))$.
Following the \textbf{Proposition 5}\,, the reduced Wigner function $W^{(1,1)}_{(1,1)}$ depending only on radius of the Bloch vector and defined  
over a 1-dimensional orbit of $SU(2)\,,$ i.e., on a circle is 
\begin{equation}
\label{eq:redWF}
 W^{(1,1)}_{(1,1)} (\theta) =\mbox{tr}\left[
    \left(
\begin{array}{cc}
 r_1 & 0  \\
 0 & r_2
\end{array}
    \right)
\Delta^{(1,1)}(\theta)
    \right]\,.
\end{equation}
The reduced SW kernel $\Delta^{(1,1)}(\theta)$ is derived from the generic kernel  (\ref{eq:SW2})  by projecting the matrix $U\in SU(2)$ written in the symmetric 3-2-3 Euler decomposition to its double coset, $U(1)\backslash SU(2)/U(1)$, 
\[
\Delta^{(1,1)}(\theta)
=\frac{1}{2}\,\exp\left(i\frac{\theta}{2}\,{\sigma}_2\right)\,
    \left(
\begin{array}{cc}
 1+\sqrt{3} & 0  \\
 0 & 1-\sqrt{3}
\end{array}
\right)\,
\exp\left(-i\frac{\theta}{2}\,{\sigma}_2\right)\,,
\]
with the  Euler angle $\theta \in [0,\pi]$ serving as  the double coset coordinate.
Evaluation of the trace in (\ref{eq:redWF}) gives  the reduced WF in the form of  dual pairing with the unistochasic matrix $B$:
\begin{equation}
\label{eq:2WFB}
 W^{(1,1)}_{(1,1)}( \theta ) = \frac{1}{2}(\boldsymbol{r}^\downarrow, B(\theta)\boldsymbol{\pi}^\downarrow) \,,  \qquad 
B (\theta) = \left(
\begin{array}{cc}
 \cos^2{\dfrac{\theta}{2}} & \sin^2{\dfrac{\theta}{2}}  \\[0.3cm]
 \sin^2{\dfrac{\theta}{2}} & \cos^2{\dfrac{\theta}{2}}
\end{array}
    \right)\,.
\end{equation}
Hence, explicitly, the reduced WF of 2-level system reads
\begin{equation}
\label{eq:reducedWF}
 W_{(1,1)}^{(1,1)}(\theta ) =
\frac{1}{2} + \dfrac{\sqrt{3}}{2}
\, (r_1-r_2)\,\cos\theta\,.
\end{equation}

\noindent{$\bullet $\,\bf  Comment on the reduced phase-space\,$\bullet $} The WF in the equation (\ref{eq:reducedWF}) is defined over a half of a unit circle. 
How can we extend it to a whole circle? 

According to the discrete symmetry of SVD decomposition, i.e., symmetry under the permutation $P$ of eigenvalues,  there are two  WF corresponding to opposite  orders, 
\begin{eqnarray}
   {}^{\downarrow}W&=& 
\frac{1}{2} + \dfrac{\sqrt{3}}{2}
\, r\,\cos\theta\,,\\
   {}^{\uparrow}W&=&
\frac{1}{2} -\dfrac{\sqrt{3}}{2}
\,r\,\cos\theta\,.
\end{eqnarray}

One can drag the permutations $P$ of eigenvalues
\(
\varrho^\prime =P\varrho P^{-1}\,
\) 
to the following transformation of the phase-space coordinate $\theta\,,$   
\[
{}^{\uparrow}W(\theta )  = ({}^{\downarrow}W(\theta ) )^P  = {}^{\downarrow}W(\theta +\pi)\,.
\]
Hence,  this relation
\[
{}^{\downarrow}W(\theta )- {}^{\uparrow}W(\theta)=\sqrt{3}\,r
\cos(\theta)\,,
\]
gives a rule to extend the domain of definition of WF to a whole circle, $\theta \in[0,2\pi]$. 

%%%%%%%%%%%%%%%%%%%% Comments on angles 
 \noindent{$\bullet $\,\bf  Comment on the reduced quasiprobability distributions and observables\,$\bullet $}
Finally, it is worth to make a comment  on a role the reduced quasiprobability distribution plays in a description of observables.

The reduced WF allows to reconstruct the spectrum of a density matrix $\varrho\,.$ 
Indeed, it is to verify that the diagonal matrix of a qutrit state can be reconstructed 
 \begin{equation}
\label{eq:ReconReducedDiag}
  \varrho_{\mathrm{diag}}= \int \mathrm{d}\theta
 \, W^{(1,1)}_{(1,1)}(\theta)\, 
\Delta^{(1,1)}(\theta)\,, 
\end{equation} 
and thus, the complete state can be  reconstructed via the SVD for density matrix $\varrho = V(\alpha^\ast, \beta^\ast)\varrho_{\mathrm{diag}}V^\dagger(\alpha^\ast, \beta^\ast)\,.$

Using the reconstruction formula (\ref{eq:ReconReducedDiag})\,, we can build the reduced symbols of operators and corresponding observables.
The  expectation value of spin-1/2 operator in the state $\varrho$\,,  
\begin{equation}
\label{eq:exp1}
    \langle \boldsymbol{S}\rangle_\varrho =\frac{1}{2}\mbox{tr}(\boldsymbol{\sigma}\varrho)=\frac{1}{2}\,\boldsymbol{r}\,,
\end{equation}
can be derived using the symbol of spin operator and WF. The symbol of spin-1/2 operator $\boldsymbol{S}=\frac{1}{2}\boldsymbol{\sigma}$ reads:
\[
W_{\boldsymbol{S}}(\Omega_2)=\mbox{tr}\left(\boldsymbol{S}
\Delta(\Omega_2\right)= \frac{\sqrt{3}}{2}\,\boldsymbol{n}\,.
\]
On the other hand, (\ref{eq:exp1}) can be  written as convolution, 
\begin{eqnarray}
\label{eq:EVconvol}
 \langle \boldsymbol{S}\rangle_\varrho &=& 
 \int \mathrm{d}\Omega_2
W_\varrho(\Omega_2)W_{\boldsymbol{S}}(\Omega_2)\\
&=&\frac{\sqrt{3}}{4}\int_{\mathbb{S}^2}\mathrm{d}\boldsymbol{n}\,[1+\sqrt{3}(\boldsymbol{n}\cdot\boldsymbol{r})]\,\boldsymbol{n}
=\frac{1}{2}\,\boldsymbol{r}\,.
\end{eqnarray}
Based on the reconstruction formula (\ref{eq:ReconReducedDiag})\,,
one can obtain the same result integrating reduced Wigner function with  spin symbol for spin operator in rotated frame,  $\boldsymbol{S}^\prime=V\boldsymbol{S}^\prime V^\dagger$:
\begin{eqnarray}
\label{eq:EVconvolReduced}
 \langle \boldsymbol{S}\rangle_\varrho &=& 
 \int_{\mathbb{S}^1} \mathrm{d}\theta\,
  W^{(1,1)}_{\boldsymbol{S}^\prime}(\theta)\,
  W^{(1,1)}{(1,1)}(\theta)\,.
\end{eqnarray}
The spin symbol is calculated with the aid of reduced SW kernel,
\[
W^{(1,1)}_{\boldsymbol{S}^\prime}(\theta)=
\mbox{tr}\left( \boldsymbol{S}^\prime\Delta^{(1,1)}(\theta)\right)\,.
\]

%%%%%%%%%%%%% QUTRIT %%%%%%%%%%%%%%%%%%%
\subsection{Wigner function of a single qutrit}
%%%%%%%%%%%%%%%%%%%%%%%%%%%%%%%%%%%%%%%%

We start with the construction of WF for a 3-level system in a mixed state using a generic 1-parametric  kernel defined over 6-dimensional symplectic manifold. Then we perform its reduction to  WF defined over 2-sphere and associated with a conventional $SU(2)$-symmetric spin-1 SW correspondence. 

\noindent{$\bullet $\,\bf  Generic qutrit state\,$\bullet $}
Assume that the qutrit is in a mixed state $\varrho \in \mathfrak{P}_3:$   
\begin{equation}
\label{eq:QutritB}
\varrho=\frac{1}{3}\,\mathbb{I} + \frac{1}{\sqrt{3}}\,\sum_{\nu=1}^{8}\xi_\nu\lambda_\nu\,.
\end{equation}
The  8-dimensional Bloch vector $\boldsymbol{\xi}$ in (\ref{eq:QutritB}) obeyы the following  constraints due to the non-negativity of the  density matrix, $\varrho \geq 0 $:  
\begin{eqnarray}
\nonumber && 0\, \leq \sum_{\nu=1}^8\,\xi_\nu\xi_\nu \leq 1\,,\\
\nonumber && 0\, \leq \sum_{\nu=1}^8\,\xi_\nu\xi_\nu -\frac{2}{\sqrt{3}}\,\sum_{\mu,\nu,\kappa =1}^8\xi_\mu\xi_\nu\xi_\kappa d_{\mu\nu\kappa}\, \leq \frac{1}{3}\,, 
\end{eqnarray}
where $d_{\mu\nu\kappa}$
 denotes the ``symmetric structure constants'' of the $\mathfrak{su}(3)$ algebra.
Equivalently,  the  mixed state $\varrho $  in 
(\ref{eq:QutritB})
can be rewritten in the SVD form as    \begin{equation}
\label{eq:3den}
    \varrho= 
  V \left(
\begin{array}{ccc}
 r_1 & 0 &  0  \\
 0 & r_2 &  0  \\
 0 & 0 & r_3
\end{array}
\right)V^\dagger\,
\end{equation}
with a unitary diagonalizing matrix   $V$  and  $SU(3)$\--invariant content of a state $\varrho $  accumulated in its ordered set of eigenvalues.
The  eigenvalues in (\ref{eq:3den}) are in one-to-one correspondence with points of the ordered  $2$-simplex,  
\begin{equation}
    \label{eq:ordered2simplex}
  \sum_{i=1}^{3} r_i = 1, \quad 1\geq r_1\geq r_2 \geq  r_{3} \geq 0\,.
\end{equation}
This simplex describes the $SU(3)$ orbit space $\mathcal{O}[\mathfrak{P}_3]$ of a qutrit.
Taking into account the unit norm condition, it is convenient to introduce two independent variables $I_3$ and $I_8$:
\begin{eqnarray}
\label{eq:specrho}
r_1=\frac{1}{3}+\frac{1}{\sqrt{3}}\,I_3+\frac{1}{3}\,I_8,
\quad
r_2=\frac{1}{3}-\frac{1}{\sqrt{3}}\,I_3+\frac{1}{3}\,I_8, \quad 
r_3=\frac{1}{3}-\frac{2}{3}\, I_8.
\end{eqnarray}
As result of this mapping, the ordered 2-simplex (\ref{eq:ordered2simplex}) in new variables $I_3$ and $I_8$ defines the following representation for orbit space $\mathcal{O}[\mathfrak{P}_3]$ of a qutrit: \begin{equation}
\label{eq:Orbitxi3xi8}
\mathcal{O}[\mathfrak{P}_3]\, \colon \ \biggl\{ I_3, I_8 \in \mathbb{R}\,\ \biggl|\,\ 0\leq I_3 \leq\frac{\sqrt{3}}{2}\,, \quad \frac{1}{\sqrt{3}}\,I_3 
\leq I_8 \leq \frac{1}{2}\biggl\} \,.  
\end{equation}
%%%%%%%% QUTRIT KERNEL
\noindent{$\bullet $\,\bf  Generic SW kernel\,$\bullet$}
For a 3-level system   (\ref{eq:condNew1})  
determine a 1-parametric family of kernels $\Delta(\Omega_3\,|\,\nu)$. 
The solutions to the equations (\ref{eq:condNew1})  are divided into classes, the \textit{generic}  and \textit{degenerate} ones. 
\begin{enumerate}
\item The spectrum of generic kernels, i.e., kernels with three different eigenvalues can be determined as follows. Fixing the decreasing order of eigenvalues and choosing a value of lowest eigenvalue  as an  independent parameter,  $\mu$, one can easily obtain from
(\ref{eq:condNew1}) the spectrum of non-degenerate SW kernel of a qutrit:  
\begin{equation}
\label{eq:OneParQutritKernel}
\mbox{spec}\left(\Delta(\Omega_3\,|\,\nu)\right)=
\left\{\frac{1-\nu+\delta}{2}\,,\,\frac{1-\nu-\delta}{2}\,, \nu\right\}
\end{equation}
with $\delta= \sqrt{(1+\nu) (5-3\nu)}\,$ and 
\(\nu \in (-1, -\frac{1}{3})\,.\)
\item Assuming a double algebraic multiplicity of eigenvalues, we convinced that the ``master equations'' 
(\ref{eq:condNew1})
admit two different solutions. Furthermore, these solutions follow from   (\ref{eq:OneParQutritKernel}) by taking limits, $\nu \to -1$ and $\nu \to -1/3$ respectively:
\footnote{The SW kernel  
(\ref{eq:DegQutrit Luis Kernels})
defines the Wigner function of a qutrit, derived by Luis in Ref.~\onlinecite{Luis2008}.}
\begin{eqnarray}
\label{eq:DegQutrit Kernels}
\mbox{spec}\left(\Delta(\Omega_3\,|\,-1)\right)&=&
\left\{\,1\,,\,1\,,\,-1\,\right\}\,, \\
\mbox{spec}\left(\Delta(\Omega_3\,|\,-\frac{1}{3})\right)&=&\left\{\,\displaystyle{\frac{5}{3}\,,\,-\frac{1}{3}\,,-\frac{1}{3}}\,\right\}\,.
\label{eq:DegQutrit Luis Kernels}
\end{eqnarray}
\end{enumerate}
In order to relate these solutions to the moduli space  of SW kernel described  in Section \ref{sec:IID}\,,  one can write SVD  (\ref{eq:SWkernelexp}) for a generic 1-parametric kernel: 
\begin{equation}\label{eq:qutritCartan}
\Delta(\Omega_3)=U(\Omega_3)\frac{1}{3}[I+ 2\sqrt{3}\left(\mu_3\lambda_3+\mu_8\lambda_8\right)]U(\Omega_3)^\dagger\,,
\end{equation}
where the  Gell-Mann basis of the $\mathfrak{su}(3)$ algebra  $\{ \lambda_1, \lambda_2, \dots,  \lambda_8\}$ (see Eq.~\ref{eq:lam-matr})  with $\lambda_3$ and $\lambda_8$ from its Cartan subalgebra is used.  Comparing decomposition (\ref{eq:qutritCartan}) with solution (\ref{eq:OneParQutritKernel}) 
one can easily  find  the coefficients $\mu_3$ and $\mu_8$ as functions of the parameter $\nu$:  
\begin{equation}\label{eq:coeffmu}
\mu_3(\nu)=\frac{\sqrt{3}}{4}\sqrt{(1+\nu)(5-3\nu)}\,, \quad 
\mu_8(\nu)=\frac{1}{4}(1-3\nu). 
\end{equation}
 A straightforward  computation shows that $\mu_3^2+\mu_8^2=1\,;$ in accordance with (\ref{eq:moduliWF}),  the moduli space of a qutrit is an arc of the unit circle. 
The corresponding polar angle $\zeta$ changes in the interval 
$\zeta \in [0,\  {\pi}/{3}]\,$ and is connected to the parameter $\nu$:   
\begin{equation}
\nonumber \nu=\frac{1}{3}-\frac{4}{3}\cos(\zeta)\,.
\end{equation}
The angle $\zeta$  serves as the moduli parameter of the unitary nonequivalent Wigner functions  of a qutrit and is related to the 3-rd order 
SU(3)-invariant  polynomial of the SW kernel:
\begin{equation}
\nonumber \det\left(\frac{1}{3}I-\Delta(\Omega_3\,|\,\nu)\right) = \frac{16}{27} \cos(3\zeta)\,,
\end{equation}
which remains ``unaffected''  by the master equation (\ref{eq:condNew1}).

%%%%%%%%%%%%% WF 
\noindent{$\bullet $\,\bf  WF of qutrit in terms of the Bloch vector\,$\bullet$} Now we pass to the derivation  of an explicit form of the Wigner function for a qutrit. With this aim the diagonalizing matrix $U(\Omega_3) \in SU(3)$ in (\ref{eq:SWkernelexp}) can be 
presented in the form of a generalized Euler decomposition
(see e.g. Refs.~\onlinecite{Byrd1998,ByrdSudarshan1998,GHKLMP2006}, and references therein)  with coordinates $\Omega_3=\{\alpha, \beta,\gamma, a, b, c, \theta, \phi \}\,, $ 
\begin{equation}
\label{eq:EulerSU3}
U(\Omega_3)=V(\alpha,\beta,\gamma)\exp\left(i\theta\lambda_5\right)V(a,b,c)\exp\left(i\phi\lambda_8\right)\,,
\end{equation}
where the left  and right factors $V$ 
denote two copies of the $SU(2)$ group embedded in $SU(3)$\,:
\begin{equation}
\nonumber V(a,b,c)=\exp\left(i\frac{a}{2}{\lambda}_3\right)\exp\left(i\frac{b}{2}{\lambda}_2\right)\exp\left(i\frac{c}{2}{\lambda}_3\right)\,.
\end{equation}
The angles in decomposition (\ref{eq:EulerSU3}) take values from the intervals  
\begin{eqnarray}
\nonumber && \alpha, a \in [0,2\pi];\,\quad  \beta, b \in [0,\pi];\,\quad 
\gamma, c \in [0,4\pi];\\ 
&&\nonumber \theta \in [0,\pi/2];\,\quad  \phi \in [0, \sqrt{3}\pi]\,.
\end{eqnarray}
These ranges allow to parameterize  almost all group elements  
(except the set of points on the group manifold whose measure is zero) and lead  to the correct value of the invariant volume of $SU(3)$ group, 
\[
\mbox{vol}(SU(3))= \int_{SU(3)}\mathrm{d}\mu_{SU(3)}=\sqrt{3}\pi^5\,.
\]

Substituting the Bloch representation for a mixed 3-level state (\ref{eq:QutritB}) 
and SW kernel decomposition  (\ref{eq:qutritCartan}) with  Euler parametrization (\ref{eq:EulerSU3}) in the expression (\ref{eq:WignerFunction}) we arrive at the following 
representations for the Wigner function of a single qutrit: 
\begin{equation}\label{eq:QutritWF1}
W^{(\nu)}_{\boldsymbol{\xi}}(\Omega_3)=\frac{1}{3}+ \frac{4}{3}\big[\mu_3\, (\boldsymbol{n}^{(3)}, \boldsymbol{\xi})+ \mu_8\, (\boldsymbol{n}^{(8)}, \boldsymbol{\xi})\big]\,,
\end{equation}
with two  orthogonal unit 8-vectors $\boldsymbol{n}^{(3)}$ 
and $\boldsymbol{n}^{(8)},$   
\begin{equation}
\nonumber n^{(3)}_\nu = \frac{1}{2}\,\mbox{tr}\left[U\lambda_3 U^\dagger 
\lambda_\nu\right], 
\; 
n^{(8)}_\nu= 
\frac{1}{2}\mbox{tr}\left[U\lambda_8 U^\dagger \lambda_\nu\right].
\end{equation}
The explicit expressions for the components of these vectors in the Euler parametrization (\ref{eq:EulerSU3}) are listed  in the Appendix ( se  Eq.~\ref{n3vec} and \ref{n8vec} respectively).

%%%%%%%%%%%%% Symmetries of WF 
\noindent{$\bullet $\,\bf Symmetries of SW kernel\,$\bullet$}
The symmetries of system set some limitation on the WF dependence on the symplectic coordinates.   
It turns out that since the regular and degenerate kernels have different isotropy groups, the corresponding diagonalizing matrices $U(\Omega_3)$ in (\ref{eq:qutritCartan}) belong to different cosets and, as a result, the WF admits a  reduction to a certain invariant subspaces of $\Omega_3\,.$ 
The symmetry types of SW kernel for 3-level system are dictated by the corresponding  isotropy groups:   
\begin{enumerate}
\item[(i).]For the regular  kernels  $H=U(1)\times U(1)\,.$
\item[(ii).]The degenerate kernel with
$\nu=-1\,$  is characterized by two equal eigenvalues of $\Delta(\Omega_3\,|\, -1)$ in the upper corner  which  means that $H=SU(2)\times U(1)$   and therefore the  Wigner function depends only on four angles:  
\begin{equation}
\nonumber W^{(-1)}_{\boldsymbol{\xi}}(\alpha,\beta,\gamma, \theta)=\frac{1}{3} + \frac{4}{3}\, (\boldsymbol{n}^{(8)}, \boldsymbol{\xi})\,.
\end{equation}
\item[(iii).]For the degenerate kernel with  
$\nu=-{1}/{3}\, $ the coefficients (\ref{eq:coeffmu}) take  the values 
$\ \mu_3 \to {\sqrt{3}}/{2}\,, \  \mu_8 \to  {1}/{2}\,$  and the Wigner function takes the form
\begin{eqnarray}\label{eq:WFp1/3in123}
 W^{(-1/3)}_{\boldsymbol{\xi}}(\alpha,\beta,
 \gamma, \theta, a,b)= \frac{1}{3} + \frac{2}{\sqrt{3}}\, 
\Big(\boldsymbol{n}^{(3)}
+ \frac{1}{\sqrt{3}}\boldsymbol{n}^{(8)}, \boldsymbol{\xi}\Big).
\end{eqnarray}
\end{enumerate}
Despite the fact that kernel $P^{(3)}(-{1}/{3})$ in (\ref{eq:DegQutrit Kernels}) has the isotropy group $H=U(1)\times SU(2)$, the Wigner function in (\ref{eq:WFp1/3in123}) shows dependence on 
six angles. This indicates that the choice
of Euler parametrization (\ref{eq:EulerSU3}) is not adapted to the isotropy group structure. To find a minimal set of four functionally independent coordinates $\{\alpha^\prime, \beta^\prime, \gamma^\prime, \theta^\prime\}$ on the coset $SU(3)/U(1)\times SU(2)$\,, 
it is necessary to consider another  embedding  of $\mathfrak{su}(2) \subset \mathfrak{su}(3)\,.$ 
Namely, using the Gell-Mann basis, we fix the subalgebra   
$\mathfrak{su}(2)=\mbox{span}
\{\lambda_6, \lambda_7, -\frac{1}{2}\lambda_3 +\frac{\sqrt{3}}{2}
\lambda_8 \}\,.$  With this choice the Euler 
decomposition for the $SU(3)$ group looks like (\ref{eq:EulerSU3}), but with the difference that both $U(2)$ subgroups are   
embedded in the ``lower corner'':  
\begin{widetext}
{\small
{\begin{eqnarray}\label{eq:Euler67}
\nonumber V(a^\prime, b^\prime, c^\prime)=
\exp\left(-i\,\frac{a^\prime}{2}\,\left(\frac{1}{2}{\lambda}_3-\frac{\sqrt{3}}{2}{\lambda}_8\right)\right)\,\exp\left(i\,\frac{b^\prime}{2}\,{\lambda}_7\right)\,
\exp\left(-i\,\frac{c^\prime}{2}\,
\left(\frac{1}{2}{\lambda}_3-\frac{\sqrt{3}}{2}{\lambda}_8)\right)\right)\,.
\end{eqnarray}
}}
\end{widetext}
As a result, the angles $a^\prime, b^\prime, c^\prime $ and $\phi^\prime$
turn out to be redundant. The Wigner function in  the newly adapted parametrization 
depends  only  on the four remaining angles through the 8-dimensional vector $\boldsymbol{n}^{\prime}$:
\begin{equation}
\label{eq:WFp1/3in123prime}
\nonumber W^{(-1/3)}_{\boldsymbol{\xi}}(\alpha^\prime,\beta^\prime,\gamma^\prime, \theta^\prime)=\frac{1}{3} + \frac{4}{{3}}\, 
(\boldsymbol{n}^{\prime}, \boldsymbol{\xi})\,. 
\end{equation}
The explicit dependence of the vector $\boldsymbol{n}^{\prime}$ on 
the angles $\{\alpha^\prime,\beta^\prime,\gamma^\prime, \theta^\prime\}$
is given by Eq.~\ref{n3vecprime}. As it was expected, the vector   $\boldsymbol{n}^{\prime}$ can be obtained from  $\boldsymbol{n}^{(8)}$ by rotation 
\begin{equation}\label{eq:nprimen8}
\nonumber\boldsymbol{n}^{\prime}\left(\alpha^{\prime}, \beta^{\prime},\gamma^{\prime},\theta^{\prime}\right)=-\boldsymbol{O}\boldsymbol{n}^{(8)}\left(\alpha, \beta,\gamma,\theta\right)\,.
\end{equation}
with the constant orthogonal  $8\times 8$ matrix 
$\boldsymbol{O}\,,$  which is the adjoint matrix $\mathrm{Ad}_{{T}}$  corresponding to the permutation ${T}$ of the first and third eigenstates of the SW kernel. Its explicit form can be found in Eq.~\ref{eq:Operm}.

\begin{figure}
    \centering
    \includegraphics[width=5 cm]{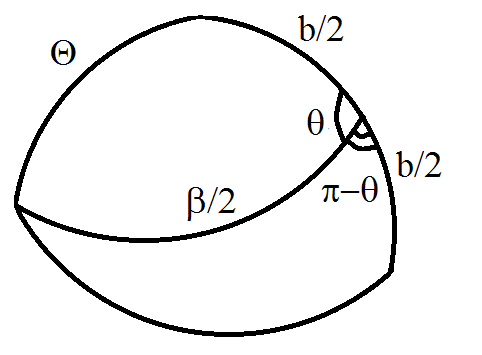}
   \caption{Geometrical meaning of the angle $\Theta$.}
    \label{spher_tri}
\end{figure}

%%%%%%%%%%%%% WF 
\noindent{$\bullet $\,\bf SW spin-1 correspondence from WF of qutrit \,$\bullet$}
Having the expression for WF of a  generic 3-level system defined  on $U(3)/U(1)^2$\,,  we are able to show how to reduce WF to the subset  $SU(2)/U(1)\,.$ The reduced Wigner  function realizes the $SU(2)$ symmetric SW  spin-1  correspondence. In construction of this SW correspondence we will proceed  similarly to the spin-1/2 case. First of all we introduce the reduced SW kernel:
\begin{equation}
\Delta^{(1,1,1)}(\boldsymbol{\chi})= Z(\boldsymbol{\chi})
\left(
\begin{array}{ccc}
 \pi_1 & 0 &  0  \\
 0 & \pi_2 &  0  \\
 0 & 0 & \pi_3 
\end{array}
\right)
Z^\dagger(\boldsymbol{\chi})\,,
 \end{equation}
where  $3\times 3$ matrix $Z(\boldsymbol{\chi})$ 
is an element of the double coset 
$S(U(1)^3)\backslash SU(3)/ S\left(U(1)^3\right)$  
\begin{equation}
    \label{eq:u3coset}
    Z(\boldsymbol{\chi})=V(0,\beta,\gamma)
    \exp\left(i\theta\lambda_5\right)V(a,b,0)\,.
\end{equation}
In (\ref{eq:u3coset}) we use the Euler representation (\ref{eq:EulerSU3})  
with  left and right factors fixed by an embedding of $SU(2)$  into the  $SU(3)$ group such that 
the five angles $\chi $ form a subset of  $\boldsymbol{\chi}= (a,b, \theta, \beta, \gamma)$  of eight Euler angles  $\{\alpha, \beta,\gamma, a, b, c, \theta, \phi \}\,, $ in (\ref{eq:EulerSU3}). Hence, the reduced 3-level WF is 
\begin{equation}
\label{eq:3WF}
    W^{(1,1,1)}_{(1,1,1)}
    (\boldsymbol{\chi}) = \mbox{tr}\left[
  \left(
\begin{array}{ccc}
 r_1 & 0 &  0  \\
 0 & r_2 &  0  \\
 0 & 0 & r_3
\end{array}
\right)   
    \Delta^{(1,1,1)}(\boldsymbol{\chi})\right]\,.
\end{equation}
Taking into account (\ref{eq:u3coset})\,, the reduced Wigner function defined in (\ref{eq:WFB1}) can be written  for 3-level system similarly to a qubit case  (\ref{eq:2WFB}) as the bilinear form 
\begin{equation}
\label{eq:3WFB}
 W^{(1,1,1)}_{(1,1,1)}(\boldsymbol{\chi}) = (\boldsymbol{r}^\downarrow, B(\boldsymbol{\chi})\boldsymbol{\pi}^\downarrow) \,,  
\end{equation}
with $3\times3 $ matrix $B(\boldsymbol{\chi}):$ 
\begin{equation}
\label{eq:3B}
B(\boldsymbol{\chi})=\left(
\begin{array}{ccc}
 B_{11} & B_{12} & \sin^2{\theta}\cos^2{\dfrac{\beta}{2}}\\[0.3cm]
% 2  
 B_{21} & B_{22}
& \sin^2{\theta}\sin^2{\dfrac{\beta}{2}}
\\[0.3cm]
\sin^2{\theta}\cos^2{\dfrac{b}{2}} &  \sin^2{\theta}\sin^2{\dfrac{b}{2}} & \cos^2{\theta}
\end{array}
    \right)\,,
\end{equation}
where elements of $2\times2 $ submatrix  are:
\begin{eqnarray}
\label{eq:B11}
B_{11}&=&\cos ^2\left(\frac{a+\gamma }{2}\right) F(\pi -\theta ,\pi -\beta ,\pi -b)^2+F(\theta ,\pi -\beta ,\pi -b)^2 \sin ^2\left(\frac{a+\gamma }{2}\right)\,,\\
\label{eq:B12}
B_{12}&=&
\cos ^2\left(\frac{a+\gamma }{2}\right) F(\theta ,\pi -\beta ,b)^2+F(\pi -\theta ,\pi -\beta ,b)^2 \sin ^2\left(\frac{a+\gamma }{2}\right)\,,\\
  \label{eq:B21}
   B_{21}&=& 
\cos ^2\left(\frac{a+\gamma }{2}\right) F(\theta ,\beta ,\pi -b)^2+F(\pi -\theta ,\beta ,\pi -b)^2 \sin ^2\left(\frac{a+\gamma }{2}\right)\,,\\
B_{22}&=&\cos ^2\left(\frac{a+\gamma }{2}\right) F(\pi -\theta ,\beta ,b)^2+F(\theta ,\beta ,b)^2 \sin ^2\left(\frac{a+\gamma }{2}\right)\,.
\label{eq:B22}
\end{eqnarray}
The function $F$ from above expressions reads:
\[
F(\theta, \beta, b) = \cos{\dfrac{\beta}{2}}
   \cos{\dfrac{b}{2}} +\cos{\theta}\sin{\dfrac{\beta}{2}}
   \sin{\dfrac{b}{2}}\,.
\]
Assuming that $\theta $ is the angle between sides $\beta/2$ and $ b/2$ of a spherical triangle, the function $F$ can be written as 
\[
F(\theta)=\cos{\Theta}\,,
\]  
where $\Theta$ is the side opposite to angle 
$\theta$  (see. Fig. ~\ref{spher_tri};
note, considering the corresponding polar triangle, the function $F(\pi-\theta, \pi-\beta,\pi-b)$ can be interpreted as a cosine of the angle opposite to the side $\theta$).
Taking into account expressions  for the 
qutrit density matrix (\ref{eq:specrho}) and 
eigenvalues of SW kernel 
\begin{eqnarray}
\label{eq:specDelta}
\pi_1=\frac{1}{3}+\frac{2}{\sqrt{3}}\,
\mu_3+\frac{2}{3}\,\mu_8,\quad
\pi_2=\frac{1}{3}-\frac{2}{\sqrt{3}}\,\mu_3+\frac{2}{3}\,\mu_8, \quad 
\pi_3=\frac{1}{3}-\frac{4}{3}\,\mu_8\,,
\end{eqnarray}
the reduced Wigner function  (\ref{eq:3WFB})  can be written as:
\begin{eqnarray}
\label{eq:3WFB2}
W^{(1,1,1)}_{(1,1,1)}(\chi) &=&\frac{1}{3} +\frac{2}{3}\,\begin{pmatrix}I_3, & I_8, \end{pmatrix}
 B^\prime
\begin{pmatrix}\mu_3, & \mu_8 \end{pmatrix}^T
 \\
 &=&\frac{1}{3}+\frac{2}{3}
 \begin{pmatrix}I_3, & I_8, \end{pmatrix}
 \left( \begin{array}{cc}
  (B_{11}+B_{22})-(B_{12}+B_{21})
& {\sqrt{3}}(B_{23}-B_{13})
\\
{\sqrt{3}}(B_{32}-B_{31}) & -(1-3B_{33})
\end{array}
    \right)
\begin{pmatrix}\mu_3 \\ \mu_8 \end{pmatrix}\,.
\nonumber
\end{eqnarray}

The $2\times 2 $ matrix $B^\prime$ in terms of Euler angles reads 
\begin{equation}
\label{eq:3WFB2}
 B^\prime
 =
\left( \begin{array}{cc}
(1+\cos^2\theta)\cos\beta\cos{b}-2\cos\theta\cos(a+\gamma)\sin\beta\sin{b}
& -{\sqrt{3}}\,\sin^2\theta\cos{\beta}
\\
 -{\sqrt{3}}\,\sin^2\theta\cos{b} & -(1-3\cos^2{\theta})
\end{array}
    \right)\,.
\end{equation}
From (\ref{eq:3WFB2}) it follows that 
WF can be rewritten as a quadratic polynomial with respect to  the variable  $
z =\cos\theta\,,$ 
\begin{equation}
\label{eq:WFgeneric}
 W= \frac{4}{3}\left\{Az^2+Bz+C\right\}\,, 
\end{equation}
where 
\begin{eqnarray}
A&=&\frac{1}{2}\left[I_3\cos(\beta) +
\sqrt{3}I_8\right]\left[\mu_3\cos(b) +\sqrt{3}\mu_8\right], \\
B&=&-I_3\mu_3\sin(\beta)\sin(b)\cos(a+\gamma),\\
C&=&\frac{1}{2}\,I_3\left[\mu_3\cos(b)-\sqrt{3}\mu_8
\right]\cos(\beta)
-
\frac{\sqrt{3}}{2}\,I_8\left[\mu_3\cos(b)+\frac{1}{\sqrt{3}}\mu_8\right]+\frac{1}{4}
\,. 
\end{eqnarray}

%%%%%%%%%%%%%SW  spin-1  correspondence
\noindent{$\bullet $\,\bf Reduction to $SU(2)$ symmetric SW  spin-1  correspondence \,$\bullet$}
According to the Table \ref{Table2}\,, the symmetry type of  SW kernel  and mixed state allowing to realize the sought reduction to $SU(2)/T^1$ is given by pairs of Young diagrams (1,1,1) and (2,1);
the necessary value of sum of squares 
is $4\times1\times2=8 $ describing SU(2) symmetric 
SW correspondence for spin-1 is: 
\[
(1,1,1)^2 + (2,1)^2 =3+5= 8 \,.
\]
\begin{enumerate}
\item {\bf SW kernel with symmetry $S(U(2)\times U(1))\,.$} The expression for the Wigner function   with ${S(U(2)\times U(1))}$ symmetric kernel follows from (\ref{eq:3WFB2}) when  $\mu_3 = 0$\,:
 \begin{equation}
    W^{(2,1)}_{(1,1,1)}(\theta, \beta)= \frac{1}{3}+\frac{2}{3}
     \left[\sqrt{3}\,(I_3\cos\beta+\sqrt{3}\,I_8)
     \cos^2{\theta}
     -(\sqrt{3}\,I_3\cos\beta+I_8)\right]\,; 
 \end{equation}
\item {\bf State with symmetry $S(U(2)\times U(1))\,.$} The expression for the Wigner function   ${S(U(2)\times U(1))}$ symmetric state follows from 
(\ref{eq:3WFB2}) when  $I_3 = 0$\,:
 \begin{equation}
 \label{eq:}
     W^{(1,1,1)}_{(2,1)}(\theta, b)= \frac{1}{3}+\frac{2}{3}\,I_8
     \left[\sqrt{3}\,(\mu_3\cos\beta + \sqrt{3}\,\mu_8)
     \cos^2{\theta}
     -(\sqrt{3}\,\mu_3\cos\beta+\mu_8)\right]\,. 
 \end{equation}
 \end{enumerate}
 %%%%%%%%%%%%%%%%%%%% Comments on angles 
 \noindent{$\bullet $\,\bf  Comments on the reduced phase-space\,$\bullet $}
 Note that the reduced WF in both cases,  1. and 2.\,, is definite on  one fourth part  of a unit sphere:
 \begin{equation}
     \label{eq:2domain}
     0\geq \theta\geq \frac{\pi}{2}\,,  
 \quad 0\geq \beta \geq \pi \,.
 \end{equation}
 According to the formulae (\ref{eq:B11})-(\ref{eq:B22})\,, the left action of  the permutation matrix $P_{12}$ on the matrix $B$
 can be dragged to the following shifts of  angles $\theta$ and  $\beta$:  
 \begin{equation}
     P_{12}B(\theta,\beta,b)= B(\theta+\pi, \beta+\pi,b)\,, \qquad  
 P_{12}=
 \begin{pmatrix}
 0 & 1 & 0\\
 1 & 0 & 0\\
 0 & 0 & 1
 \end{pmatrix}\,.
 \end{equation}
 Therefore, the domain of definition of angles in (\ref{eq:2domain}) can be extended  to cover a whole unit 2-sphere.  

%%%%%%%%%%%%%%%%%%%%%
\section{Concluding remarks}
\label{sec:VConclusion}
%%%%%%%%%%%%%%%%%%%%%%%%

In the present article we argue the existence of the unitary non-equivalent representations for the Stratonovich-Weyl kernels corresponding to the Wigner functions of arbitrary $N$\-- dimensional quantum system. 
The admissible Wigner functions  can  be classified by the values of 
$SU(n)$\--invariant polynomials in the elements of the SW kernel. As it was shown, the ``master equation'' (\ref{eq:condNew1}) fixes the values only of the lowest degree polynomial invariants, the first and second ones,  while  values of the   remaining $N-2$ algebraically independent invariants distinguish members of the family of SW kernels.

In conclusion, it is necessary to mention that the present consideration  of the quasiprobability functions makes no difference between  elementary and composite systems. In forthcoming publications we will discuss in detail the Wigner functions for  composite quantum systems, paying special attention to the manifestations of  ``quantumness'', particularly the entanglement
\cite{BengtssonZyczkowskiBOOK}, in properties of quasidistributions. 

Apart from this, we leave  for  future  studies   interrelations  between the reduction of WF described in the present note for a system possessing  symmetries, with the corresponding Hamiltonian reduction on the phase-space of its classical counterpart.

\appendix*
\section{}
\subsection{The Gell-Mann basis of $\mathfrak{su}(3)$}
%%%%%%%%%%%%%%%%%%%%%%%%%%%%%%%%%%%%%%%%%%%%%
The Gell-Mann basis $\{\lambda_1, \lambda_2,\dots,\lambda_8\}$ of the Lie algebra
$\mathfrak{su}(3)$ reads:
\begin{equation}\label{eq:lam-matr}
\begin{array}{c}
\begin{array}{ccc}
 \lambda _{1}=\left(\begin{array}{ccc}
   0 & 1 & 0 \\
   1 & 0 & 0 \\
   0 & 0 & 0 \
 \end{array}\right)\,,&
 \lambda _{2}=\left(\begin{array}{ccc}
   0 & -i & 0 \\
   i & 0 & 0 \\
   0 & 0 & 0 \
 \end{array}\right)\,,& \\ & & \\
  \lambda _{3}=\left(\begin{array}{ccc}
   1 & 0 & 0 \\
   0 & -1 & 0 \\
   0 & 0 & 0 \
 \end{array}\right)\,, &
  \lambda _{4}=\left(\begin{array}{ccc}
   0 & 0 & 1 \\
   0 & 0 & 0 \\
   1 & 0 & 0 \
 \end{array}\right)\,,& \\ & &\\  
 \lambda _{5}=\left(\begin{array}{ccc}
   0 & 0 & -i \\
   0 & 0 & 0 \\
   i & 0 & 0 \
 \end{array}\right)\,,&
  \lambda _{6}=\left(\begin{array}{ccc}
   0 & 0 & 0 \\
   0 & 0 & 1 \\
   0 & 1 & 0 \
 \end{array}\right)\,,\
 \end{array}
 \\ \\
\begin{array}{cc}
\lambda _{7}=\left(\begin{array}{ccc}
   0 & 0 & 0 \\
   0 & 0 & -i \\
   0 & i & 0 \
 \end{array}\right)\,,&
  \lambda _{8}=\displaystyle{\frac{1}{\sqrt{3}}}\left(
  \begin{array}{ccc}
   1 & 0 & 0 \\
   0 & 1 & 0 \\
   0 & 0 & -2 \
\end{array}\right)\,.
\end{array}\
\end{array}
\end{equation}
%%%%%%%%%%%%%%%%%%
\subsection{The adjoint action of the permutation matrix $T$}

Consider the matrix which permutes the first and third diagonal elements of the $3\times 3$ identity matrix: 
\begin{equation}\label{eq:perm}
 T = \left(\begin{array}{ccc}
0 & 0 & 1\\
0 & 1 & 0\\
1 & 0 & 0
\end{array}
\right)\,.
\end{equation}
The adjoint matrix $\mathrm{Ad}_{T}$ corresponding to the permutation (\ref{eq:perm}),  
$
T \lambda_\mu T = \left(\mathrm{Ad}_{{T}}\right)_{\mu\nu}\lambda_\nu\,,
$ is   
\begin{equation}\label{eq:Operm}
\mathrm{Ad}_{T}= \left(
 \begin{array}{cccc|cccc}
0 & 0 & 0     & 0 & 0 & 1 &0  &  0        \\
0 & 0 & 0     & 0 & 0 & 0 &-1  &  0         \\
0 & 0 & 1/2  & 0 & 0 & 0 & 0 & -\sqrt{3}/2 \\
0 & 0 & 0     & 1 & 0 & 0 & 0 &  0         \\
\cline{1-8}
0 & 0 & 0     & 0 & -1 & 0 & 0 &  0         \\
1 & 0 & 0     & 0 & 0 & 0 & 0 &  0         \\
0 & -1 & 0     & 0 & 0 & 0 & 0 &  0         \\
0 & 0 & -\sqrt{3}/2   & 0 & 0 & 0 & 0 & -1/2 
\end{array}
\right)\,.
\end{equation}

\subsection{The adjoint vectors of SU(3)}

Using the Euler decomposition (\ref{eq:EulerSU3}), we determine the adjoint matrix $\mathrm{Ad}_{U}$ of SU(3) 
transformations $U$:  
\begin{equation}\label{eq:AdjRepr}
U \lambda_i U^\dagger =\left(\mathrm{Ad}_{U}\right)_{ij} \lambda_j\,, \qquad
\mathrm{Ad}_{U}  \in SO(8)\,.
\end{equation}
Below, only expressions for vectors  $n^{(3)}_i=(\mathrm{Ad}_{U})_{3i}$ and $n^{(8)}_i=(\mathrm{Ad}_{U})_{8i}$, specifying  the Wigner function  of a single  qutrit   (\ref{eq:QutritWF1}), will be presented. 
Components of the vector $\boldsymbol{n}^{(8)}$ read:
%\vspace{0.5cm}
\begin{widetext}
\begin{equation}\label{n3vec}
\begin{array}{ccl}
 n^{(3)}_1&=&\Big(\sin(\alpha)\sin(a+\gamma)-\cos(\alpha)\cos(\beta)\cos(a+\gamma)\Big)\sin(b)\cos(\theta) 
+\cos(\alpha)\sin(\beta)\cos(b)
\Big(1-\frac{1}{2}\sin^2(\theta)\Big),\\&&\\
%%%  2
 n^{(3)}_2&=&\Big(\cos(\alpha)\sin(a+\gamma)+
\sin(\alpha)\cos(\beta)\cos(a+\gamma)
\Big)\sin(b)\cos(\theta)
+\sin(\alpha)\sin(\beta)\cos(b)\Big(1 -\frac{1}{2}\sin^2(\theta)\Big),\\&&\\
%%%  3
 n^{(3)}_3&=& -\cos (a+\gamma)\sin(\beta)\sin(b)\cos(\theta) + 
\cos(\beta)\cos(b)\Big(1-\frac{1}{2}\sin^2(\theta)\Big), \\&&\\
%%% 4
 n^{(3)}_4&=&\cos\left(\frac{\alpha-\gamma}{2}-a\right)\sin\left(\frac{\beta}{2}\right)\sin(b)\sin(\theta )-
\frac{1}{2}\cos\left(\frac{\alpha +\gamma }{2}\right)\cos\left(\frac{\beta}{2}\right)
\cos(b)\sin(2\theta),\\&&\\
%%% 5
 n^{(3)}_5&=&\sin\left(\frac{\alpha-\gamma}{2}-a\right)\sin\left(\frac{\beta }{2}\right)\sin(b)\sin(\theta)+
\frac{1}{2}\sin\left(\frac{\alpha+\gamma}{2}\right)\cos\left(\frac{\beta}{2}\right)\cos(b)
\sin (2\theta),\\&&\\
%%% 6
n^{(3)}_6&=&\cos\left(\frac{\alpha+\gamma}{2}+a\right)
\cos\left(\frac{\beta}{2}\right)\sin (b)\sin(\theta)+
\frac{1}{2}\cos\left(\frac{\alpha-\gamma}{2}\right)\sin\left(\frac{\beta}{2}\right)\cos(b)\sin(2\theta),\\&&\\
%%% 7
 n^{(3)}_7&=&\sin\left(\frac{\alpha+\gamma}{2}+a\right)
\cos\left(\frac{\beta }{2}\right)\sin (b)\sin (\theta)
+\frac{1}{2}\sin \left(\frac{\alpha-\gamma}{2}\right)
\sin\left(\frac{\beta }{2}\right)\cos(b)
\sin(2\theta),\\&&\\
%\end{array}
%\end{equation}
%\begin{equation}
%%% 8
%\nonumber \hspace{-12.5cm} 
n^{(3)}_8 &=&-\frac{\sqrt{3}}{2}\cos(b)\sin^2(\theta).
\end{array}
\end{equation}

The 8-vector $\boldsymbol{n}^{(8)}$ depends only on four angles 
$\{\alpha, \beta, \gamma, \theta \}$ and its components are:t4   
\begin{equation}\label{n8vec}
\begin{array}{ccl}
%%%% 1
n^{(8)}_1&=&+\frac{\sqrt{3}}{2}\cos(\alpha)\sin(\beta)\sin^2(\theta), \\ && \\
%%% 2
n^{(8)}_2&=&-\frac{\sqrt{3}}{2}\sin(\alpha)\sin(\beta)\sin^2(\theta),\\ &&\\
%%%% 3
n^{(8)}_3&=&-\frac{\sqrt{3}}{2}\cos(\beta)\sin^2(\theta),\\&&\\
%%% 4 
n^{(8)}_4&=&-\frac{\sqrt{3}}{2}\cos\left(\frac{\alpha +\gamma}{2}\right)\cos\left(\frac{\beta}{2}\right)
\sin(2\theta),\\&&\\
%%% 5
n^{(8)}_5&=&+\frac{\sqrt{3}}{2}\sin\left(\frac{\alpha+\gamma }{2}\right)
\cos\left(\frac{\beta}{2}\right)\sin(2\theta),\\&&\\
%%% 6
n^{(8)}_6&=&+\frac{\sqrt{3}}{2}\cos\left(\frac{\alpha -\gamma}{2}\right)
\sin\left(\frac{\beta}{2}\right)\sin(2\theta),\\&&\\
%%% 7
n^{(8)}_7&=&+\frac{\sqrt{3}}{2}\sin \left(\frac{\alpha -\gamma}{2}\right)
\sin\left(\frac{\beta}{2}\right)\sin (2\theta),\\&&\\
%%% 8
n^{(8)}_8&=&1-\frac{3}{2}\sin^2(\theta).
\end{array}
\end{equation}
The 8-dimensional vector $\boldsymbol{n}^{\prime}$ in formula 
(\ref{eq:WFp1/3in123prime})
reads 
\begin{equation}\label{n3vecprime}
\begin{array}{ccl}
 n_{1}^{\prime}&=&-\frac{\sqrt{3}}{2}\cos \left(\frac{\alpha^{\prime}-\gamma^{\prime}}{2}\right)\sin \left(\frac{\beta^{\prime}}{2}\right)\sin(2\theta^{\prime})\,,  \\&&\\
%%%%% 2
  n_{2}^{\prime}&=&-\frac{\sqrt{3}}{2} 
\sin\left(\frac{\alpha^{\prime}-\gamma^{\prime}}{3}\right)\sin \left(\frac{\beta^{\prime}}{2}\right)\sin(2\theta^{\prime})\,,  \\&&\\
%%%%% 3
 n_{3}^{\prime}&=&\frac{\sqrt{3}}{2}
\left[\cos^2(\theta^\prime) 
-\sin^2\left(\frac{\beta^{\prime}}{2}\right)
\sin^2(\theta^{\prime})\right]\,,\\&&\\
%%%%% 4
n_{4}^{\prime}&=&-\frac{\sqrt{3}}{2} 
\cos\left(\frac{\alpha^{\prime}+\gamma^{\prime}}{2}\right)  
\cos\left(\frac{\beta^{\prime}}{2}\right)\sin(2\theta^{\prime})\,,  \\&&\\
%%%%% 5
n_{5}^{\prime}&=&\frac{\sqrt{3}}{2} 
 \sin\left(\frac{\alpha^{\prime}+\gamma^{\prime}}{2}\right)\cos\left(\frac{\beta^{\prime}}{2}\right)\sin(2\theta^{\prime})\,,  \\&&\\
%%%%% 6
 n_{6}^{\prime}&=&\frac{\sqrt{3}}{2}\cos(\alpha^{\prime})
 \sin(\beta^{\prime})\sin^2(\theta^{\prime})\,, \\&&\\
 %%%%% 7
 n_{7}^{\prime}&=&-\frac{\sqrt{3}}{2}  \sin (\alpha^{\prime} ) \sin(\beta^{\prime})\sin^2(\theta^{\prime})\,, \\&&\\
 %%%%% 8
 n_{8}^{\prime}&=&\frac{1}{2} \left[1-3 \cos ^2\left(\frac{\beta^{\prime} }{2}\right) \sin ^2(\theta^{\prime})\right]\,.
\end{array}
\end{equation}
\end{widetext}

%%%%%%%%%%%%%%%%%%%%%%%%%%%%%%%%%%%%%%%%%%%%%%%%%%%%%5
%\section*{References}


\begin{thebibliography}{19}
\bibitem{footnote} For example, in the quatrit case, $N=4\,,$  any fixed order of eigenvalues corresponds to one out of 24 possible ways to tessellate a sphere by the spherical triangles whose angles are  
$(\pi/2\,,{\pi}/{3}\,,{\pi}/{3})$. Such a triangle is one of the four fundamental spherical M\"obius  Triangles with the tetrahedral symmetry, which is  
classified as a (2,3,3) triangle.
Repeated reflections in the sides of the triangles will tile a sphere exactly once. In accordance with Girard's theorem, the spherical excess of a triangle determines the solid angle: $\pi/2 + \pi/3 + \pi/3 - \pi = 4\pi/24\,.$
\end{thebibliography}
\end{document}